\documentstyle[twocolumn,aps]{revtex}
\begin{document}

\title{Recent Developments in the Theory of Scarring}
\author{L. Kaplan\thanks{kaplan@physics.harvard.edu}
 \\Department of Physics and Society of Fellows,\\ Harvard
University, Cambridge, MA 02138}
\maketitle


\begin{abstract}

We review recent progress in attaining a quantitative
understanding of the scarring phenomenon, the non-random behavior of
quantum wavefunctions near unstable periodic
orbits of a classically chaotic system.
The wavepacket dynamics framework
leads to predictions about statistical long-time
and stationary properties of quantum systems with chaotic classical analogues.
Many long-time quantum properties can be quantitatively understood using only
short-time classical dynamics information; these
include wavefunction intensity distributions,
intensity correlations in phase space and correlations between
wavefunctions, and distributions of decay rates and conductance
peaks in weakly open systems. Strong deviations from random matrix theory are
predicted and observed in the presence of short unstable periodic orbits.

\end{abstract}

\section{Introduction}

Wavefunction scarring is the anomalous enhancement of
quantum eigenstate intensities along unstable
periodic orbits of a classically
chaotic system; it is surely one of the most visually striking properties
of quantum chaotic behavior. Observed numerically in unpublished work
by McDonald~\cite{mcdonald}, scars were later brought to the attention
of the physics community by Heller~\cite{hellerscar}, who also provided
the first theoretical explanation for their existence. Numerical 
evidence and associated analytical work (followed later by
experimental tests in a variety of systems) showed that scarring was
a statistically significant correction to Berry's early
conjecture~\cite{berryconj} that
eigenstates of a classically ergodic system
are evenly distributed over the energy hypersurface, in the semiclassical
limit $\hbar \to 0$. 

The structure of quantum eigenstates of a system whose classical analogue
is ergodic has long been a matter of interest for physicists and
mathematicians alike. When an integrable system is quantized, the 
resulting eigenstates can be well understood in terms of the WKB quantization
condition along the classical periodic orbits~\cite{bt}.
For classically ergodic
systems, such a direct correspondence between quantum wavefunctions
and classical periodic orbits is, however, not possible. Because the typical
classical trajectory evenly
explores all of the available energy hypersurface, it
is natural to suppose that the quantum eigenstates also have constant 
intensity over this entire hypersurface, up to the inevitable
quantum fluctuations. Rigorous results have been obtained on
one important aspect of this problem. Theorems by Schnirelman,
Zelditch, and Colin de Verdiere 
\cite{scdvz} consider a classically defined operator and show that
its expectation value over almost all individual eigenstates converges
(in the semiclassical limit)
to the ergodic, microcanonical average of the classical version of
the operator. In this limit, however,
taking the expectation value means averaging over more and more
de Broglie wavelengths of the wavefunction, since the classical
symbol of the operator in phase space is assumed to be kept fixed
as $\hbar \to 0$. Thus, these ergodicity theorems provide little information
about the structure of eigenstates at the single-wavelength scale (or more
generally on scales of a single momentum-space channel or over cells
of size O($\hbar$) in phase space).

Classical ergodicity by itself is certainly not sufficient to ensure
ergodically distributed quantum wavefunctions over individual wavelengths
or channels.
Thus, for example, one can consider the ``slow ergodic systems,"~\cite{wqe}
where the classical exploration of phase space is sufficiently slow that 
the corresponding quantum dynamics can only explore a fraction of the
available quantum channels by the Heisenberg time [this being
$\hbar$ over the mean
quantum level spacing, the time at which the quantum exploration of
phase space effectively ends and the quantum dynamics becomes
quasi-periodic]. In such systems, the participation ratio, i.e. the
fraction of available
quantum channels with which a typical eigenstate has significant overlap,
approaches zero in the semiclassical limit $\hbar \to 0$. However, for
any given high-energy eigenstate, this small fraction of bright channels
are evenly distributed over the total phase-space, so that eigenstate
intensity integrated over a classically-defined patch of phase space
does in fact converge to the classically required value, in perfect
accordance with the results of Schnirelman, Zelditch, and Colin de
Verdiere (SZCdV).

Such anomalies are not known to arise for strongly
chaotic systems, by which we mean
systems that have everywhere positive entropy. In such systems, the mixing
time (the time for a phase space cell of size Planck's constant
$h$ to explore all of the
available phase space on a mesh of that scale) scales only logarithmically 
with $h$, in contrast with the Heisenberg time, which scales as a
power law. Thus, in the limit of small $h$ a given classical system
will always have a mixing time much shorter than the Heisenberg time, and
many trajectories will exist linking any $h$-sized cell with any other
well before individual eigenstates begin to be resolved. It is reasonable to
conjecture~\cite{bgs} that for such systems individual eigenstates
will in fact evenly cover all of the available energy hypersurface,
up to uncorrelated
Gaussian fluctuations given by random matrix theory (RMT).
Indeed, RMT
does correctly predict much of the structure of eigenstates (and also
the eigenvalue spectrum) of classically chaotic systems.

However, we also
know that the Gutzwiller trace formula~\cite{gutz} is quite
effective in using
short periodic orbits to predict
non-random fluctuations in the quantum spectra of classically
chaotic systems. It is then
not too surprising that the unstable periodic orbits of a chaotic system
should also have a strong effect on the {\it eigenstate} structure, modifying
RMT predictions. In fact, it appears that in the $\hbar \to 0$
limit, scarring constitutes the most important deviation from perfect
ergodicity of chaotic eigenstates at the $\hbar$-scale, in the sense of RMT.
Other deviations from RMT, such as dynamical localization, disappear in the
semiclassical limit, but the scarring phenomenon survives. It, along with
symmetry effects, is sufficient to explain quantitatively many properties 
of quantum chaotic systems, including wavefunction intensity statistics,
wavefunction and phase space intensity correlations,
channel-to-channel transport,
resonance widths, and conductances.

We conclude this introduction by defining what we mean by scarring and
addressing four common misconceptions about this phenomenon;
our comments here may help to clarify issues which in the past
have led to confusion and even controversy surrounding this subject.

(i) Scarring is defined as the anomalous enhancement or suppression of
eigenstate intensity on or near an {\it unstable} periodic orbit and its
invariant manifolds\footnote{The
invariant manifolds are the set of
phase space points that classically approach the periodic orbit as
$t \to +\infty$ (the {\it stable} manifold), and the set that approach the
same orbit as $t \to -\infty$ (the {\it unstable} manifold). In a 
two-dimensional phase space,
these are both curves, and the lines tangent to the
curves at the periodic point are known as the stable and unstable
directions, respectively. The (infinitely many) points of intersection of the
stable and unstable manifolds are the
{\it homoclinic} points of the orbit (see below).}, in a chaotic system.
Of course, stable periodic orbits
also attract eigenstate intensity, as mentioned above,
but the reasons for this are well understood in terms of the semiclassical
theory of integrable systems~\cite{bt}. More importantly, the increased
tendency of a probability distribution launched in the vicinity of a stable
orbit to overlap with itself at long times can be understood as a purely
classical phenomenon. Phases are necessary only to obtain the quantization
energies, in accordance with the WKB condition. On the other hand scarring
can be thought of in the time domain as the increased probability of a quantum
wavepacket launched near an {\it unstable} orbit to overlap with itself
at very long times. Classical dynamics in a chaotic system shows no
such behavior: classically an evolved probability distribution loses 
at long times all
memory of its initial position. So scarring is at first sight
a paradoxical phenomenon because its presence implies that quantum
mechanics retains a much better memory at long times of short-time
classical behavior than does long-time classical mechanics itself. This
paradox, as we will see, is resolved by realizing that scarring is inherently
an interference effect, and can only be understood by including phase
information.

(ii) Scarring is a {\it statistically significant} deviation from
single-channel quantum ergodicity, as defined by random matrix theory (which
predicts uniform wavefunction amplitude over all of available
phase space, up to
Gaussian random fluctuations). Of course random enhancements in
wavefunction intensity occur even in the absence of any periodic orbit; they
can be observed, for example, in random superpositions of plane waves
of constant energy~\cite{scarlet}. Thus, anecdotal evidence, in the form
of pictures of scarred states, is generally not a good way to test
a theory of
scarring, because even in RMT there is a finite probability to observe
any given enhancement in one or a few states. An exception to this general
rule arises when the instability exponent $\lambda$ of a given
periodic orbit is very small. Then as we shall see one expects to find
many states (more precisely, a fraction of $O(\lambda)$ of all states)
having enhancement factors of order $1/\lambda$ at this orbit,
and each one of these
individually has an exponentially small probability to occur within the context
of RMT. For moderate $\lambda$, a few states will not
produce the desired evidence of scarring, and one needs to consider the
entire distribution of eigenstate intensities on the periodic orbit, sweeping
through energy or some other parameter. This distribution can then be compared
quantitatively with the scar theory prediction (which differs very
significantly 
from RMT).

(iii) Scarring in not the same as the semiclassical evaluation of
individual eigenstate intensities; rather it is a statistical
property of states near a periodic orbit. It is true that in certain systems
semiclassical methods work well all the way to the Heisenberg time, and can
be used to resolve individual eigenstates~\cite{ltsc}. However,
this requires putting in at least as much information into the semiclassical
calculation as there is in the full quantum system itself.
Exact wavefunctions of
individual states are sensitive to small changes in system
parameters, and they provide no intuition about the overall
behavior of the system. On
the other hand, the statistical predictions
of scar theory (concerning the distribution of wavefunction intensities, mean
conductances and conductance fluctuations, mean resonance lifetimes, etc.)
depend
on only a few parameters (e.g., the exponent $\lambda$), and provide
intuitively useful information about entire ensembles of chaotic systems which
could not be obtained by using a computer to diagonalize a given Hamiltonian.
The distinction is also very important experimentally because the
scarring phenomenon is robust under small perturbations to the system;
thus scar theory predictions will often be valid in situations where
the exact dynamics is not known well enough to allow individual eigenstates
to be theoretically determined (either quantum mechanically or
semiclassically).

(iv) Finally, as explained above, scars are not a threat to SZCdV
ergodicity; scar theory addresses quantum structure at the $\hbar$-level
and is to be properly compared with RMT, a much stronger condition
than the quantum ergodicity of Schnirelman, Zelditch, and
Colin de Verdiere.

The remainder of this paper is organized as follows: in the next section we 
briefly review some relevant experiments and discuss several theoretical
approaches that have been used to understand scarring. Because the literature
is extensive, it will not be possible to do full justice to the contributions
of the various authors. The bulk of the present paper 
is a survey of results recently obtained using the wavepacket dynamics
method, a project initiated by Heller in the seminal paper\cite{hellerscar}.
Of course, many or all of the results presented here can be understood
also in the context of the other approaches, particularly in the context
of Gutzwiller
periodic orbit theory. However, as a unifying framework we choose to use
the time-domain methodology in which the results were first obtained, and make
contact with other points of view where appropriate.

Thus, Section~\ref{lnls} introduces and reviews the linear (short-time)
and nonlinear (long-time) theory of scarring, as discussed in greater detail in
\cite{nlscar}. The emphasis here is on developing the conceptual framework
and describing the relations among the 
concepts of the Gaussian wavepacket, the short-time and
long-time autocorrelation function, the local density of states (LDOS), the
linearized energy envelope, and the inverse participation ratio (IPR).
This machinery is put into use immediately in Section~\ref{qt} where
we perform numerical tests and compare the data with quantitative
theoretical predictions about spectra, IPR's, and energy correlations.
An analysis of wavefunction intensity statistics~\cite{sscar}
follows in Section~\ref{wis}, where we find in accordance
with scar theory a power-law tail in the intensity distribution, in sharp
contrast with the exponential-tail prediction of RMT. Section~\ref{ims}
addresses and resolves ambiguities in the measure of scarring intensity, 
leading to a universal and fully optimal measure in the strong scarring
limit\cite{is}. Section~\ref{open} deals with more recent applications of scar
theory to open systems, discussing resonances, the probability to remain
in an open chaotic system\cite{open}, and the effect of scars on
conductances\cite{cond}.
Finally, in Section~\ref{conc} we sum up the results and
suggest some directions for the future.

\section{Experimental evidence and theoretical progress}

\subsection{Experimental and numerical evidence for scars}

Scarred eigenfunctions were first directly observed experimentally
in two-dimensional microwave cavities by Sridhar~\cite{sridhar},
and independently by Stein and St\"ockman~\cite{stockman}.
Here one makes use of the
exact correspondence between the quantum wavefunctions of a billiard system
and the electromagnetic modes in a cavity with the same
boundary conditions. The electromagnetic modes at various energies
can be mapped either by measuring the reflected microwave power as a function
of energy and position~\cite{stockman}, or by using a perturbative technique
involving measurements of small energy shifts~\cite{sridhar}.
Individual eigenstates scarred
by specific short unstable orbits can be observed, although no attempt was
made to quantify the scarring strengths. Of course, experimental
limitations may preclude the collection of data at extremely high frequencies,
where one could test scar theory predictions about the statistics
of chaotic wavefunctions in the semiclassical limit.

Another very important set of experiments~\cite{fromhold} have been performed
in semiconductor heterostructures. A resonant tunneling device contains
a quantum well separated by a barrier from a two-dimensional electron
gas. Electrons can be injected into the quantum well, and information about
electron wavefunctions obtained by studying the tunneling rates. When
a large tilted magnetic field is turned on, the classical dynamics of
the system is known to be chaotic. Indeed, very strongly scarred wavefunctions
are observed in such systems, associated with orbits that have a small
instability exponent for a wide range of system parameters~\cite{ns}.
Again, the scarring patterns are in good qualitative agreement with theoretical
expectations.

Given the difficulty in collecting large amounts of good experimental
data at very high energies, numerical simulations of quantum chaotic
systems have from the very beginning been useful in providing
``empirical" support for the scarring
phenomenon. Many different quantum chaotic
systems have been thus explored, including, for example,
the hydrogen atom in a uniform magnetic field~\cite{wintgen}. In recent
work by Li and Hu~\cite{li}, a plane wave decomposition method is used
to find wavefunctions near the one-millionth state of the stadium billiard,
showing that strong scarring is still present at
such energies. Certain orbits, including the diamond orbit, are observed
to be scarred in a corridor of width scaling as the wavelength (rather
than as the geometric
mean of the wavelength and the system size; see below).
The reason for this is not well
understood by the present author, though it may be associated with the
very special properties of the stadium, namely, the presence of marginally
stable orbits and intermittency.

A statistical analysis of intensities in highly-excited
wavefunctions of the stadium and other hard chaotic systems
would prove very useful; such a detailed analysis of medium-energy
wavefunctions in phase-space coordinates on the billiard
boundary is now in preparation\cite{bies}. A somewhat different
point of view is taken by B\"acker, Schubert, and Stifter~\cite{backer},
who examine the approach to SZCdV quantum ergodicity in position space for
various chaotic billiards (again, in a medium energy range).
Simonotti, Vergini, and Saraceno~\cite{saraceno} have studied scar intensity 
on the boundary of a stadium billiard, using a phase-space representation
and also considering improved
test states which are superpositions of {\it several}
phase-space Gaussians (see Section~\ref{ims}).

\subsection{Several theoretical approaches}

The first theoretical explanation of the scarring phenomenon is found already
in \cite{hellerscar}, where a Husimi space measure of scar strength is used.
In this approach (which we discuss at much greater length beginning with 
the following section), wavefunction intensity in phase space is defined
by the squared overlaps of the wavefunction with Gaussian wavepackets
$|a^{q_0,p_0}\rangle$
centered at all possible phase space points $(q_0,p_0)$.
If we consider one such Gaussian
and look at its squared overlaps with all eigenstates of the system, 
we obtain
a local density of states at the location of that wavepacket
\begin{equation}
\label{sq0p0}
S^{q_0,p_0}(E)=\sum_n \left |\langle n|a^{q_0,p_0} \rangle\right 
|^2 \delta(E-E_n) \,.
\end{equation}
What Heller realized was that the autocorrelation function of the wavepacket,
\begin{equation}
\label{at}
A^{q_0,p_0}(t)= \langle a^{q_0,p_0} | a^{q_0,p_0}(t) \rangle \,,
\end{equation}
the Fourier transform of the spectrum $S(E)$,
shows large
recurrences at short
times which are integer multiples of the orbit period $T_P$, if
the point $(q_0,p_0)$ is on or near a periodic orbit of the system. The
strength of these
recurrences falls off exponentially with time, roughly as
$e^{-\lambda t/T_P}$, where $\lambda$ is the instability exponent of
the unstable periodic orbit. In the semiclassical limit $\hbar \to 0$,
these periodic
recurrences (both amplitudes and phases)
can be easily analytically computed in terms of the classical stability
matrix and the classical action
of the periodic orbit. The recurrences are called ``linear",
as they are obtained by linearizing the dynamics around the unstable periodic
orbit in question. This short-time structure in $A(t)$ produces an
envelope in the spectrum $S(E)$ which must be the smoothed version of the
true LDOS at the periodic orbit. Thus, by considering
only the short-time dynamics of the Gaussian wavepacket, we obtain
an energy-smoothed LDOS,
\begin{equation}
S^{q_0,p_0}_{\rm smooth}(E)=
\sum_n |\langle n|a^{q_0,p_0} \rangle|^2 w(E-E_n)
\end{equation}
from which we can easily
extract the average wavefunction intensity near various energies. If 
the exponent $\lambda$ is small, the smoothed LDOS is
shown to have a sequence of bumps separated by $2\pi\hbar/T_P$ and of width
scaling as $\lambda \hbar/T_P$. At these optimal scarring energies
(which are analogous to the EBK energies for an integrable system),
wavefunctions must, on average, have larger than normal intensities on
the periodic orbit, while far from these spectral peaks wavefunctions
must be typically antiscarred, i.e. have less than expected intensity
on the same orbit.

The linear theory of scarring provides information only about the average
wavefunction intensity in some energy range; it does not tell us whether
most of that intensity is contained in a few wavefunctions (the 
``totalitarian" scenario) or whether it is equally distributed among
all the wavefunctions near that energy (the ``egalitarian" scenario).
For this, we need a nonlinear theory of scarring, one that takes into
account long-time recurrences in the autocorrelation function $A(t)$.
These recurrences correspond to pieces of the wavepacket leaving the vicinity
of the periodic orbit, traveling through other regions of phase
space where they are subjected to complicated chaotic dynamics, and eventually
returning to the periodic orbit. In many situations, these long-time
recurrences may be treated statistically, producing a separation
of scales between the classically-obtained short-time behavior near
the periodic orbit and the random long-time behavior, as suggested
already by Antonsen et al.~\cite{ott}. What is important to note here
is that the short-time behavior must necessarily leave an imprint on
the long-time behavior and on stationary properties like the LDOS,
as we will see more clearly in the following section.

First, we briefly mention some alternative theoretical approaches to
the scarring problem, which provide useful and complementary perspectives
on the subject. A theory of wavefunction
scarring in position space has been developed by
Bogomolny~\cite{bogo}; clearly the position space basis is a very
relevant one for many experimental applications. Bogomolny again smoothes
the wavefunction intensity over some small energy range $\Delta E$,
\begin{equation}
<|\psi(q)|^2 >_{\Delta E} = {1 \over N} \sum_n |\psi_n(q)|^2 \,,
\end{equation}
and (in the semiclassical limit $\hbar \to 0$)
represents this smoothed intensity as a sum over periodic trajectories
of the system:
\begin{eqnarray}
\label{bogosum}
& & <|\psi(q)|^2 >_{\Delta E} =
\rho_0(q)+\hbar^{(n-1)/2} \nonumber \\ \times \sum_p {\rm Im}
& & \left< B_p(z)\exp  \left( i {S_p \over \hbar} + 
i { W_p^{km}(z) \over 2\hbar}
y_k y_m\right)  \right>_{\Delta E} \,.
\end{eqnarray}
For each periodic orbit $p$, $z$ is chosen to be a coordinate along the
orbit, while the $y_m$ ($m=2\ldots n$)
are all the coordinates perpendicular to the orbit. $S_p$
is the classical action around the orbit, while $B_p$ is a focusing factor
which can be easily obtained from the stability matrix of the orbit (i.e.
from the linearized dynamics for small $y,\dot y$). Similarly $W_p^{km}$
measures the quadratic change in the action for small values of $y$, as
one moves away from the orbit. It is important to
note here that the sum in Eq.~\ref{bogosum} is effectively
finite: any orbits of period
$T_p >> \hbar / \Delta E$ can be dropped as their oscillatory contribution
vanishes when integrated over the smoothing interval $\Delta E$. As
$\Delta E$ becomes small, more and more orbits need to be included
in the sum (the number of orbits grows exponentially with period $T_p$
in a chaotic system). The constant $\rho_0(q)$ term can be thought of
as coming from the zero-time dynamics of the system; it is related to the mean
density of states at energy $E$.
Bogomolny computes the oscillatory
contributions to the above sum from various 
short periodic orbits in the stadium billiard and finds qualitative agreement
with numerical data. 

Bogomolny's semiclassical Green's function approach is clearly closely
related to our wavepacket dynamics method, as the semiclassical Green's
function can be obtained from the semiclassical time-domain propagator
by a stationary-phase Fourier transform. One difference between the approaches
is that Bogomolny envisions summing over a large number of periodic
orbits to get as close as possible to an energy domain resolution of order
of a mean level spacing. As mentioned above, in some systems it is
indeed possible to use semiclassical methods to compute individual
eigenstates of the system\cite{ltsc}.
In fact for this purpose one needs information only about 
orbits of period up to the mixing time (which scales logarithmically
with $\hbar$) rather than the Heisenberg time (which scales as a power law).
However, our aim here is to make predictions about the distribution of
scarring
strengths based only on linearized information around {\it one} periodic
orbit; for this purpose most other orbits which produce additional
oscillations in the density of states may be treated
statistically~\cite{ott}. It is important to note in this context that
if we are measuring wavefunction intensities on a given orbit $p$
of period $T_p$, then in the semiclassical limit there are no short orbits 
that come close to this orbit (on a scale of $\hbar$)
in phase space. The only oscillatory
contributions which will need to be taken into account are from orbits 
closely related to orbits {\it homoclinic} to $p$ [homoclinic orbits
are those that approach $p$ at large negative times, perform an excursion
away from $p$ into other regions of phase space, and then
again approach $p$ at large positive times]. In fact, in the $\hbar \to 0$
limit the periodic orbit sum  (Eq.~\ref{bogosum}) for a point $q$
on a given periodic orbit $p$ can be written equivalently as a contribution
from orbit $p$ itself plus a sum over trajectories homoclinic to $p$. Although
the two points of view are mathematically equivalent, the homoclinic
sum approach makes explicit the special role of the orbit $p$
near which we are making measurements.

We also take this opportunity to note that a position space basis, though
obviously physically natural in many measurement situations, is not
generally an optimal one for detecting scar effects. Unless the
periodic point $q$ happens also to be a focusing point of classical
trajectories near the orbit, only a small fraction of the total scar
strength is captured in the position basis, and the fraction becomes smaller
as $\hbar$ decreases (or as the energy increases).
An easy way to see this is to notice that the effects
of a classical trajectory in quantum mechanics {\it generically}
extend to a region around
the orbit scaling not as a wavelength but rather as the square root of a
wavelength (and similarly the affected region scales as the square root
of the total number of channels in momentum space).
Thus, unless either the stable or unstable manifold of
the orbit $p$ at periodic point $q$ happens to be oriented along the
momentum direction, the position space basis will not be optimal, as
reflected in the falling off of the focusing prefactor $B_p$
with energy (and similarly the momentum basis will not be optimal, unless
one of the two invariant manifolds is oriented along the position direction).
All this will become more clear in the exposition
of the following section. In any case, one should keep in mind that a
position space basis can always be considered as a special limiting case
of the Gaussian wavepacket test state, where the position uncertainty
of the wavepacket becomes comparable to a wavelength, and the momentum
uncertainty becomes large.

Wigner phase space analysis of the scarring phenomenon was
pioneered by Berry~\cite{berryscar}. Berry considers the Wigner spectral
function, $W(x,E,\Delta E)$, again smoothed over an energy interval $\Delta E$
near $E$. $x=(q_0,p_0)$ is a phase space point. Being formulated
in phase space, the approach
more closely resembles that of Heller, and working in Wigner phase
space instead of Husimi space also eliminates the need to choose the
(apparently arbitrary) eccentricity and orientation of the Gaussian
wavepackets $a^{q_0,p_0}$. The downside of Wigner phase space is the
absence of a positivity condition on $W$; thus the value of the spectral
function cannot be considered as corresponding to an intensity or a
probability of being found near a certain point $x$ (and random matrix
theory is therefore not applicable). The Husimi
function, which is manifestly positive definite,
can be thought of as a phase space smoothing of the Wigner
distribution over a phase space region scaling as $\hbar$. The ambiguity in
choosing the Gaussian centered on $x$ over which this smoothing
is to be performed is indeed an important issue, to be considered
carefully towards the end of the following section.
There we will see that to obtain
the {\it maximal} scarring effect, the Gaussian must be chosen to be properly
oriented along the stable and unstable directions at the periodic point.
[An arbitrarily large wavepacket width is allowed along either of these
directions, with a correspondingly small width in the
orthogonal direction. Also, strong, but non-maximal,
scarring will generally be obtained
for any wavepacket with width scaling as $\sqrt\hbar$ in both the
position and momentum directions.]
See also
Section~\ref{ims} for a discussion of even better measures of scarring
possible if one is willing to go beyond the Gaussian wavepacket
approximation.

Berry studies carefully the structure of the spectral Wigner function $W$
as $x$ moves off the closed orbit within the energy hypersurface and
also as $x$ moves off the energy surface itself (the fringes he obtains
disappear upon Husimi averaging). In particular, as $x$ moves
off the periodic orbit, the transverse
oscillations take the form of a complex Gaussian, with fringe spacing
scaling as $\sqrt\hbar$. The amplitude of the fringes depends only on the 
stability matrix, although the precise pattern changes as one moves along
the orbit. In a two-dimensional phase space, the fringes form
hyperbolas approaching the stable and unstable manifolds of the orbit.

A common limitation of the analyses ~\cite{hellerscar,bogo,berryscar}
is that they make no predictions about the properties of the
spectral fluctuations on scales much smaller than $\hbar/T_D$, where
$T_D \sim T_P/\lambda$
is the decay time of the unstable orbit being studied. Therefore is it not
possible to make quantitative predictions about individual wavefunction
intensities, participation ratios, etc., without explicitly doing
a Gutzwiller sum over {\it all} periodic orbits. Even if the sum can
be performed, it is by no means clear that it will converge in all
cases (e.g. in systems where caustics are important~\cite{caustic}).
When the sum does converge it may produce individual semiclassical
wavefunctions very different from the quantum eigenstates, due to
diffraction and other ``hard quantum" effects. Furthermore, as we discussed
in the introduction, such Heisenberg-time calculations are extremely
sensitive to small perturbations on the system. What one would like
is to be able to say precisely how often a given single-wavefunction
scar strength will appear
on a given orbit, at what energy, and at what parameter values. In the
semiclassical limit, this can in fact be done using only information
about linearized dynamics near the orbit itself, and, in some cases,
about a few strong isolated homoclinic recurrences which cannot
be treated statistically.

Before proceeding to the main part of the paper, we mention some of
the important contributions to scar theory since the early work
of Heller, Bogomolny, and Berry. Many different aspects of the
scarring
phenomenon have been theoretically investigated, and we here
mention only a few which are most relevant to the present work.
Agam and Fishman~\cite{fishman} define the weight of a scar by integrating
the Wigner function over a narrow tube in phase space,
of cross-section $\hbar$, surrounding the periodic orbit.
The Fredholm method
can be used to obtain a
semiclassical formula for scars. Alternatively, de Polavieja, Borondo, and
Benito~\cite{borondo} construct a test state highly localized on a given
periodic orbit using short-time quantum dynamics. We defer
all discussion of improved measures of scarring beyond the simple Gaussian
wavepacket measure until Section~\ref{ims}, where this important
issue is treated in full generality.

Klakow and Smilansky~\cite{smilansky} have used a scattering approach to
quantization to study the wavefunctions of billiard systems. They
treat carefully the wavefunctions on the Poincare surface of section,
and relate their properties to scarring in configuration space.
de Almeida~\cite{dealmeida} uses the Weyl representation to establish
connections between classical and quantum dynamics, with
particular application to the semiclassical Wigner function and scars.
Tomsovic~\cite{tomsovic} has used parametric variation as a new method
for studying scar effects; scars are shown to induce correlations
between wavefunction intensities on a periodic orbit and the level
velocities of these wavefunctions when certain system parameters are varied.
We also mention the work of
Arranz, Borondo, and Benito~\cite{arranz} who have probed the intermediate
region between regular and strongly chaotic quantum behavior,
and have shown how scarred states first arise from the mixing of pairs
of regular wavefunctions as $\hbar$ is decreased (but well before one
reaches the semiclassical limit which is the main focus of the present
review). Finally, several groups~\cite{voros,sarbaker,octbaker}
have studied the hyperbolic scar structures associated not only with the
periodic orbit itself but with its invariant manifolds and homoclinic orbits.
This subject we also return to in Section~\ref{ims}.

\section{Linear and nonlinear theory of scarring}
\label{lnls}

Consider an arbitrary (unstable) periodic orbit of a chaotic system. We will
simplify the initial exposition by taking the orbit to be a fixed point of a
discrete-time area-preserving map on a two-dimensional phase space. [The
case of a higher-period orbit of such a map, or of an orbit of
a continuous-time dynamics in two spatial dimensions, can be 
reduced to the present case simply by iterating the original map, or by taking
a surface of section map, respectively. The generalization to these two
situations is straightforward, and is discussed towards the end of this
section and in more detail in~\cite{nlscar}. Similarly, the entire
framework generalizes with minor modifications to unstable orbits in
higher-dimensional systems.] 

The fixed point is taken to be at the origin of phase space, $(q,p)=(0,0)$.
Without loss of generality, we can take the stable and unstable directions
at the fixed point to be vertical ($p$) and horizontal ($q$),
respectively. The local dynamics around the periodic orbit can always
be put into this form via a canonical transformation of the coordinates.
Then the only parameter describing the local (linearized) dynamics near the
orbit is $\lambda$, the instability exponent for one iteration of the orbit.
Locally, the equations of motion are given by
\begin{eqnarray}
\label{eqmo}
q \to \tilde q &=&e^{\lambda t}q \nonumber \\
p \to \tilde p&=&e^{-\lambda t}p \,.
\end{eqnarray}
We are interested in studying the behavior of the quantum wavefunctions
near this fixed point; as discussed earlier the natural way to do
this is by examining their overlaps with a phase-space Gaussian
centered on this point:
\begin{equation}
\label{wavepkt}
a_\sigma(q)=\left({1 \over \pi \sigma^2 \hbar}\right)^{1/4}
 e^{-q^2/2\sigma^2\hbar} \,.
\end{equation}
This is a minimum-uncertainty state centered at the origin of phase space,
with width $\sigma\sqrt\hbar$ in
the $q-$direction and $\sqrt\hbar/\sigma$
in the $p-$direction. The width $\sigma$ is an arbitrary parameter:
$\sigma^2$ is the aspect ratio of the
phase-space Gaussian, typically chosen to be of order unity. Any real
value of $\sigma$, as we shall soon see, produces a
wavepacket capable of optimally measuring scar strength.
The apparent ambiguity in the
choice of $\sigma$ is an important issue that we will return to in
Section~\ref{ims}.

In fact, there is another, less obvious, ambiguity in the choice of the test
wavepacket in Eq.~\ref{wavepkt}. A non-circular Gaussian can be made
to have any desired orientation in the $q-p$ plane by complexifying 
$\sigma$; this produces a test state in which $q$ and $p$ are
correlated. Any Gaussian which is not oriented along the stable
and unstable axes ($p$ and $q$, in our example) will be non-optimal
in the sense of having smaller short-time recurrences in its autocorrelation
function $A(t)$ compared with an optimal wavepacket, and therefore having
also smaller fluctuations in its spectral intensities
$|\langle n|a \rangle|^2$. A
wavepacket defined to be circular
in a coordinate system in which the two invariant directions
are {\it not} orthogonal will be transformed into such a ``tilted" wavepacket
in our preferred coordinate system (Eq.~\ref{eqmo}) defined by the
stable and unstable directions. Also, a position
or momentum state will generically appear ``tilted" in this preferred
coordinate system of the periodic orbit.
See Eq.~\ref{alinq} below, and the associated discussion.

For small enough $\hbar$, the wavepacket $|a_\sigma\rangle$ and its
short-time iterates are contained well within the linear regime.
As long as the wavepacket stays in the phase space region surrounding
the periodic orbit in which the linearized
equations of motion (Eq.~\ref{eqmo}) apply,
the evolution of the wavepacket is completely semiclassical, given simply
by the stretching of the $q-$width parameter $\sigma$ (and the
associated shrinking of the momentum-width $\sigma_p=\sigma^{-1}$).
More explicitly,
at short times we have 
\begin{equation}
\label{wpevol}
U^t |a_\sigma\rangle \approx
U_{\rm lin}^t |a_\sigma\rangle = e^{-i\phi t} |a_{\sigma e^{\lambda t}}
\rangle \,,
\end{equation}
where $U$ is the unitary operator implementing the quantum discrete-time
dynamics, $U_{\rm lin}$ represents the quantization of the
linearized behavior near
the periodic orbit, and $t$ is time, measured in units of a single mapping.
Here
$-\phi$ is a phase associated with one iteration of the periodic orbit:
it is given by the classical action in units of $\hbar$, plus
Maslov indices if appropriate.

Then the short-time autocorrelation function of the wavepacket is easily seen
from Eqs.~\ref{wavepkt},\ref{wpevol} to have the form
\begin{equation}
\label{shortcorr}
A_{\rm lin}(t) = e^{-i\phi t} \langle a_\sigma | a_{\sigma e^{\lambda t}}
\rangle
= {e^{-i\phi t} \over \sqrt{\cosh(\lambda t)}} \,.
\end{equation}
The `lin' subscript indicates that
Eq.~\ref{shortcorr} describes the piece of the autocorrelation function
coming from the linearized dynamics around the periodic orbit.
For a weakly
unstable orbit (small $\lambda$), $A_{\rm lin}(t)$
is slowly decaying, with strong
recurrences happening for the first $O(1/\lambda)$ iterations of the orbit.
We note that the short-time autocorrelation function $A_{\rm lin}(t)$
is $\sigma$-independent (for real $\sigma$),
a fact that will prove important later on.

After a certain time, the wavepacket leaves the linearizable region and
nonlinear recurrences
begin to dominate the return probability. This time scale, called
the log-time, scales logarithmically in $N \sim \hbar^{-1}$ because of the
exponential divergence of trajectories away from the orbit:
\begin{equation}
\label{logtime}
T_{\rm log} \sim {\log fN \over \lambda}\,.
\end{equation}
Here $N$ is the total number of
$h$-sized cells in the accessible phase space (also equal to the dimension
of the effective quantum mechanical Hilbert space), and $f$ is the fraction
of this phase space (typically $O(1)$) in which the linearized equations
of motion (Eq.~\ref{eqmo}) apply.
Semiclassically, the long-time recurrences are given by a sum
over all trajectories homoclinic to the original periodic orbit. These
are classical trajectories that as $t \to -\infty$ approach the fixed
point along the unstable manifold (the $q$-axis, in our case), and
as $t \to +\infty$ approach it again, this time along the
stable manifold ($p$). Let us choose a point $x=(\delta q,0)$
on the unstable axis which is well within the linearizable region
surrounding $(0,0)$, and which after $T$ time steps maps
to a point $f^T(x)=(0,\delta p)$ on the stable axis, again inside the
linearizable regime. In between the trajectory
must undergo some complicated
nonlinear behavior in phase-space regions far from the fixed point.

The path of the full trajectory and its surroundings
naturally separates into three stages: (i) $\tau_1$ steps
during which a thin vertical strip centered on $x$
of the original wavepacket shrinks
vertically and stretches horizontally as  its center moves out horizontally
at an exponential rate along the unstable manifold, finally reaching the
edge\footnote{Of course, the exact boundaries of the linearizable region
are arbitrary and in no way affect the final result. All that is required is
for the wavepacket itself to be well-contained in the region where the
linearized equations of motion hold.}
of the linearizable region; followed
(ii) by complicated nonlinear dynamics for $T-\tau_1-\tau_3$ steps,
which eventually brings the center of
the distribution back into the linearizable region; and finally
(iii) a time of $\tau_3$ steps during which the dynamics takes us
back through the linearizable region, this time along the stable
direction.
After $T$ steps, the original thin vertical strip has become a long
horizontal strip which then intersects the original Gaussian. The total
contribution to the wavepacket autocorrelation function at time $T$
coming from this homoclinic excursion is given by a product of five
factors:
\begin{eqnarray}
\label{hc}
A_{\cal HC} & = &
e^{-\delta q^2/2\sigma^2} \cdot e^{-i\tau_1\phi}e^{-\lambda\tau_1/2}
\nonumber \\ & \cdot &  Q(T-\tau_1-\tau_3)e^{i\phi_{\rm nonlin}}
\nonumber \\ & \cdot &
e^{-i\tau_3\phi}e^{-\lambda\tau_3/2} \cdot
e^{-\delta p^2/2\sigma_p^2} \,.
\end{eqnarray}
The factors $e^{-\lambda\tau_1/2}$ and $e^{-\lambda\tau_3/2}$ are
instability factors associated with the linearized motion of the wavepacket,
while $e^{-i\tau_1\phi}$ and $e^{-i\tau_3\phi}$ are the corresponding phases.
The suppression factors $e^{- \delta q^2/2 \hbar \sigma^2}$
and $e^{- \delta p^2 /2\hbar \sigma_p^2}$
result from the fact that the initial and final points of the
excursion are both off-center relative to the Gaussian wavepacket
(recall $\sigma_p=1/\sigma$).
The total correlation function $A(T)$ is given semiclassically by a sum
of terms of the form (Eq.~\ref{hc}) over all the homoclinic excursions
of length $T$:
\begin{equation}
\label{hcsum}
A_{\rm SC}(T)=\sum_{\cal HC} A_{\cal
HC} \,.
\end{equation}

Because the long-time homoclinic orbits come back with complicated accumulated
phases $\phi_{\rm nonlin}$, 
and the number of these recurrences grows exponentially with time,
one might expect the total long-time return amplitudes to be given by Gaussian
random numbers.
In fact, however, we must consider together contributions from
all homoclinic excursions which lie on a single homoclinic orbit,
e.g. the trajectory which takes $x \to f^T(x)$ in $T$ steps and
the one which takes 
$f^{\Delta_1}(x) \to f^{T+\Delta_2}(x)$
in $T+\Delta_2-\Delta_1$ steps, where $\Delta_1$ and $\Delta_2$ are small
enough that the resulting phase-space points are still in the linearizable
regime.
These paths all come back in phase with each other, having the same nonlinear
phase $\phi_{\rm nonlin}$, and give
rise to short-time correlations in $A(t)$ for large $t$ \cite{nlscar}.
These correlations are shown to be
related to the short-time dynamics of the original
Gaussian wavepacket. In fact,
we can write the return amplitude at times $T_{\rm log} \ll T \ll T_H$
($T_H=N$ is the Heisenberg time, where individual eigenstates
begin to be resolved) as
a convolution
\begin{equation}
\label{convol}
A(T)=\sum_\tau A_{\rm rnd}(\tau) A_{\rm lin}(T-\tau) \,.
\end{equation}
Here $A_{\rm lin}$ is the short-time return amplitude, and $A_{\rm rnd}$
has the statistical properties of an uncorrelated random Gaussian
variable. In particular,
\begin{eqnarray}
\label{rnd}
<A_{\rm rnd}(\tau)> & = & 0 \nonumber \\
<A^\star_{\rm rnd}(\tau)A_{\rm rnd}(\tau')> & = &
 {1 \over N} \delta_{\tau \tau'}\,.
\end{eqnarray}
The prefactor $1/N$ provides the proper classical normalization: in the absence
of interference effects, the probability to come back is equal to the
probability for visiting any other state in the Hilbert space.
The average in Eq.~\ref{rnd} is taken over long times $\tau$,
$T_{\rm log} \ll \tau \ll T_H$, and/or over an ensemble of systems
which all have the same linearized dynamics around our chosen periodic orbit.
In either case, the total size of the Hilbert space $N$ ($=1/h$ for a phase
space area normalized to unity) has been assumed to
be large.
We then obtain
\begin{eqnarray}
\label{longcorr}
<A(t)> & = & 0 \nonumber \\
<A^\star(t)A(t+\Delta)> & = & {1 \over N} \sum_s
A^\star_{\rm lin}(s) A_{\rm lin}(s+\Delta)  \,.
\end{eqnarray}
At times beyond the Heisenberg time, this gets modified \cite{nlscar} to
\begin{equation}
\label{longcorr2}
<A^\star(t)A(t+\Delta)> = {F \over N} \sum_s
A^\star_{\rm lin}(s) A_{\rm lin}(s+\Delta)  \,.
\end{equation}
$F$ is a factor associated with the discreteness of the eigenstates: it is
$3$ for real eigenstate--test state overlaps and $2$ for complex overlaps.

The long-time autocorrelation function is thus self-correlated on a scale
$\Delta \sim \lambda^{-1}$. Qualitatively, this can be understood on a purely
classical level: once probability happens to come back to the vicinity of a
weakly unstable periodic orbit, it tends to stay around before leaving again.
On the other hand, the overall enhancement in the total return probability
at long times:
\begin{equation}
\label{enhprob}
<|A(t)|^2> = {F \over N} \sum_s {1\over \cosh(\lambda s)} \,,
\end{equation}
obtained by combining the general expression Eq.~\ref{longcorr2} with
the short-time overlap dynamics of the Gaussian wavepacket
(Eq.~\ref{shortcorr}), is fundamentally an interference phenomenon, and signals
a kind of quantum localization. In terms of homoclinic orbits, the enhancement
arises because of the coherent addition of paths of the same length $T$ lying
on the same homoclinic orbit (setting
$\Delta_1=\Delta_2$ in the discussion following
Eq.~\ref{hcsum}).

We now define $S(E)$ to be the Fourier transform of the autocorrelation
function,
\begin{equation}
\label{spectrum}
S(E)={1 \over 2\pi} \sum_{t=-\infty}^{+\infty} A(t) e^{iEt}\,.
\end{equation}
For a non-degenerate spectrum, it is easy to see (by inserting a complete set
of eigenstates) that this produces the line spectrum of Eq.~\ref{sq0p0}.
Cutting off the sum in Eq.~\ref{spectrum} at $\pm T_{\rm log}$,
or equivalently by including only linearized dynamics around
the periodic orbit, we obtain {\it the smoothed local density of states}:
\begin{equation}
\label{smspectrum}
S_{\rm lin}(E)=\sum_t A_{\rm lin}(t) e^{iEt}\,,
\end{equation}
an envelope centered
at quasienergy $E=\phi$ (see Eq.~\ref{shortcorr}),
 of width $\delta E \sim \lambda$, and of height
$\sim \lambda^{-1}$ (a factor of $2 \pi$ has been inserted into the
definition of $S_{\rm lin}$ for future
convenience). $E=\phi$ is the analogue of the EBK quantization condition for
integrable systems; here, because of the instability of the orbit, scarred
states can live in an energy range of $O(\lambda)$ around the optimal energy.
States with energy more than $O(\lambda |\log \lambda|)$
away from resonance tend to
be {\it antiscarred}, i.e. they have less than expected intensity at the
periodic orbit [see Eq.~\ref{inters} and Fig.~\ref{figdecslin}].

Now long-time (nonlinear) recurrences as in Eq.~\ref{convol} lead to
fluctuations under the short-time
envelope in the full spectrum $S(E)$. Because these
recurrences involve a random variable {\it convoluted}
with the short time dynamics,
in the energy domain we obtain random fluctuations {\it multiplying}
the short-time envelope. (It is easy to see physically that the random
oscillations must multiply the smooth envelope: if they were merely added to
it, the total spectrum would go negative away from the peak of the envelope.)
Finally, at the Heisenberg time $T_H = N$,
individual states are resolved~\cite{nlscar,sscar}, and we see a line spectrum
(Eq.~\ref{sq0p0}) with a height distribution given by
\begin{equation}
\label{ran}
I_{n a_\sigma} \equiv
|\langle n|a_\sigma\rangle|^2 = r_{an} S_{\rm lin}(E_n) \,.
\end{equation}
Here $r_{an}$ are random variables (with mean $<r_{an}>={1/N}$)
drawn from a chi-squared distribution
of one degree of freedom (two degrees of freedom for complex
$\langle n|a_\sigma\rangle$). Thus we obtain a random
(Porter-Thomas) line spectrum, multiplying the original linear
envelope [see Fig.~\ref{fignumspec} for a numerical example].

We now need to discuss the concept of an
inverse participation ratio (IPR),
a very useful measure for studying deviations
from quantum ergodicity. We define
\begin{equation}
{\rm IPR}_{a_\sigma} = N \sum_n I_{n a_\sigma}^2 =
N \sum_n |\langle n|a_\sigma\rangle|^4 \,.
\end{equation}
(Note that $\sum_n I_{n a_\sigma}=1$ by normalization.)
Being the first non-trivial moment of the eigenstate intensity
($I_{n a_\sigma}$)
distribution, the IPR provides a convenient one-number measure
of the strength of scarring (or any other kind of deviation from
quantum ergodicity). The IPR
would be unity for a wavepacket that had equal overlaps with all the
eigenstates of the system; the maximum value of $N$ is reached in
the opposite (completely localized regime), when the wavepacket is itself
a single eigenstate. Random matrix theory predicts an IPR of $F$, the
strong quantum ergodicity factor defined above in Eq.~\ref{longcorr2}.

However, from Eqs.~\ref{sq0p0},\ref{spectrum} we see that
\begin{equation}
\label{iprsum}
{\rm IPR}_{a_\sigma} = \lim_{T \to \infty}
{N \over T} \sum_{t=0}^{T-1} |A(t)|^2 \,;
\end{equation}
as one might expect, localization is associated with an enhanced return
probability at long times. Now from Eq.~\ref{enhprob} we see that
scar theory predicts an enhancement in the IPR over random matrix theory:
\begin{equation}
\label{scaripr}
{\rm IPR}_{a_\sigma} = F \sum_s {1\over \cosh(\lambda s)} 
\to F {\pi \over \lambda} \,,
\end{equation}
where the limit of small $\lambda$ has been taken.
($F$, as before, is $3$ or $2$,
depending on whether the states are real or complex, respectively.)
The IPR thus decomposes into a product of two
contributions: the shape of the short-time envelope coming from linear
dynamics around the periodic orbit, and a quantum fluctuation factor $F$,
as predicted by Porter-Thomas statistics. See Fig.~\ref{figiprvslin}
in the following section for numerical evidence in support of this prediction.

Before concluding this
section, we briefly mention several straightforward
generalizations of the above analysis. Eq.~\ref{shortcorr} for the short-time
autocorrelation function $A_{\rm lin}$ and the
linear spectral envelope $S_{\rm lin}$ which follows from it apply to a fixed
point of a discrete-time map, but the analysis extends easily to
higher-period orbits and to dynamics in continuous time. For a period
$T_P$ orbit of a map, the linearized autocorrelation function of
an optimally oriented wavepacket takes
the form
\begin{equation}
\label{shorthp}
A_{\rm lin}(t) 
= {e^{-i\phi t/T_P} \over \sqrt{\cosh(\lambda t/T_P)}} \delta_{t \,
\rm mod \, T_P,0} \,,
\end{equation}
where $\lambda$ is now of course the instability exponent for one iteration
of the {\it entire} orbit, and $\phi$ the corresponding phase.
This result is the same on each of the periodic
points of the orbit; however, the optimal Gaussian wavepacket will in general
have a different orientation at each point,
due to the rotation of the invariant manifolds
as one moves along the orbit. The resulting smoothed local density
of states $S_{\rm lin}$ then has $T_P$ peaks in the quasienergy interval
$[0,2\pi)$, each with a height scaling as $\lambda^{-1}$ as before
(in the regime of small instability $\lambda \ll 1$),
and a width scaling as $\lambda/T_P$. Notice that the time and quasienergy
coordinates have been simply rescaled by $T_P$; the ratio of scar peak
separation to peak width, as well as the inverse participation ratio,
remain independent of the period $T_P$, both
scaling simply as $\lambda^{-1}$.

For a continuous-time Hamiltonian dynamics, the spectral envelope has
infinitely many peaks, with energy spacing $\hbar/T_P$, height
scaling as $\lambda^{-1}$ as before, and width scaling as
$\hbar \lambda/T_P$ for small $\lambda$. The positions of the peaks 
correspond of course to the EBK quantization condition along a stable
orbit, and the widths can be thought of as resonance widths due
to the instability of the orbit. In the semiclassical limit, more
and more states are found under each of the scarring peaks, and the
individual line heights are distributed in accordance with
Eq.~\ref{ran} (see \cite{bies} for a numerical example in the stadium
billiard). The continuous-time dynamics also contains an additional
time scale not found in the discrete-time system, namely the time
$T_{\rm free}$ for the test wavepacket to traverse itself near $t=0$ and again
every time it returns to the vicinity of its original location. Thus,
the $\delta$-function sum in Eq.~\ref{shorthp} must, for a continuous-time
dynamics, be convoluted with a very short-time window of width $T_{\rm free}$
($T_{\rm free} \ll T_P$ in the semiclassical limit).
In the energy domain, this
results in the infinite sequence of scarring peaks being multiplied
by a very broad envelope of width $E_{\rm free} = \hbar/T_{\rm free}$.
$E_{\rm free}$ is
the energy uncertainty of the test wavepacket $a_\sigma$
($E_{\rm free} \sim p \sqrt\hbar \sigma_p = p \sqrt\hbar/\sigma_z$),
and evidently depends on $\sigma_z$,
the position-uncertainty of the wavepacket in units of $\sqrt\hbar$ along
the direction of the orbit.
This broad energy envelope
is a classical effect (i.e. a classical phase-space distribution corresponding
to the quantum wavepacket has this distribution of energies),
and must first be taken into account before deviations from quantum ergodicity,
such as scarring, can be analyzed. See \cite{nlscar,is,ergodic} for a more
complete discussion of these issues.

The individual peak height distribution (Eq.~\ref{ran}) in the local density
of states applies in the semiclassical limit for a hard chaotic system.
In this limit, the homoclinic (nonlinear) recurrences are all weak,
and no
individual recurrence can affect significantly either a particular
eigenstate or statistical wavefunction properties such as the IPR. However,
at finite $\hbar$ it is quite possible for a certain class of identifiable
homoclinic recurrences at medium times
$T_{\rm med}$ to have a significant effect on
the spectrum. The contributions from these special recurrences can
then be computed analytically, and an improved spectral envelope created which
takes into account not only the periodic orbit dynamics on scale
$T_P/\lambda$, but also the special strong recurrences at the time scale
$T_{\rm med}$. The general procedure, discussed in more detail in \cite{nlscar},
is a sequential analysis of the various time scales, beginning with
the shortest times. Thus, we start with the classical envelope coming
from the free dynamics on time scale $T_{\rm free}$,
then we include the effects
of the short periodic orbit on which our wavepacket is located, then
the effect of any special class of strong homoclinic recurrences
(an $\hbar$-dependent contribution), and finally the effects of random 
recurrences at long times, all the way to the Heisenberg time $T_H$.
In \cite{nlscar} an example is given of a system where the phase-space
mixing away from the periodic orbit being studied is slow, so that
at finite $\hbar$ very distinct oscillations in the LDOS at the orbit 
are obtained which can be assigned to a special class of intermediate-time
recurrences.

The expression (Eq.~\ref{shortcorr}) is valid, as we have discussed,
for any wavepacket optimally oriented along the invariant manifolds
of the periodic orbit. There is a one-parameter family of such optimal
wavepackets at each periodic point: in the coordinate system of Eq.~\ref{eqmo}
in which the stable and unstable directions are orthogonal, this free
parameter is the width of the Gaussian along either of the two directions.
Gaussians that are tilted relative the invariant directions are non-optimal:
the degree of non-optimality can be described by a single parameter $Q \ge 0$.
This is easy to see by transforming to a coordinate system in which the chosen
Gaussian is circular: in this coordinate system, $Q$ is a function of the 
angle between the stable and unstable directions ($Q=0$ for a right angle and
increases as the angle decreases to $0$). In the coordinate system in which
the manifolds are orthogonal, $Q$ will be a function of the
aspect ratio of the Gaussian and its tilt relative to the axes.
The general form of the short-time autocorrelation function is then
\begin{equation}
\label{alinq}
A_{\rm lin}(t) =
 \langle a | a(t) \rangle = {e^{-i \phi t} \over \sqrt{\cosh{\lambda t}
+iQ\sinh{\lambda t}}} \,.
\end{equation}
As long as $Q$ is not large, the qualitative scarring behavior is unchanged,
although spectral oscillations become less pronounced and there
is less scarring and antiscarring as $Q$ increases. Analytic results
are less easy to obtain for non-zero $Q$, but 
quantitative predictions can be readily produced for comparison
with any experimental or numerical data. See \cite{open} for a discussion
of the effect that non-zero $Q$ has on antiscar strength, and thus
on the probability to remain in a weakly open system.

Finally, we conclude this section by mentioning that the analysis can
be easily extended to test-states centered off of the periodic orbit.
In the coordinate system defined by Eq.~\ref{eqmo}, we may take a
Gaussian centered at $(q_0,p_0)$, with width $\sigma \sqrt\hbar$
in position space and $\sigma_p \sqrt\hbar$ in momentum (where
$\sigma_p=1/\sigma$):
\begin{equation}
\label{wavepktoff}
a_\sigma(q)=\left({1 \over \pi \sigma^2 \hbar}\right)^{1/4}
 e^{-(q-q_0)^2/2\sigma^2\hbar+ip_0(q-q_0)} \,.
\end{equation}
Then the autocorrelation function is given by \cite{genforms}
\begin{eqnarray}
\label{alinoff}
A_{\rm lin}(t) & =& {\exp{\left [-{\sinh^2{\lambda t/2} \over \cosh{\lambda t}}
(q_0^2/\sigma^2\hbar+p_0^2/\sigma_p^2\hbar)\right]}
\over \sqrt{\cosh \lambda t}}
\nonumber \\ & \times & e^{-i\phi t-i(\tanh{\lambda t})q_0p_0/\hbar} \,.
\end{eqnarray}
We easily see that the linear recurrences, and thus
scarring, are strong in a region of size $\sigma \sqrt\hbar \times
\sigma_p \sqrt\hbar = \hbar$ in phase space. Eq.~\ref{alinoff} will
be made use of in Section~\ref{wis}, where the tail of the wavefunction
intensity distribution is obtained, summing contributions from test states
located everywhere in phase space.

\section{Quantitative tests}
\label{qt}

We now provide numerical evidence supporting some of the
conclusions of Section~\ref{lnls}. The data is taken from ~\cite{nlscar},
where the reader will find additional numerical tests along with a
discussion of the generalized baker maps, the test systems used to obtain
numerical data
in this and the following section. The baker maps are a class
of Bernoulli systems and a paradigm of hard chaotic behavior. A symbolic
dynamics allows one to easily identify the locations of the periodic orbits
and to find their exponents. Furthermore, the local dynamics of these
systems around any periodic orbit
is exceedingly simple, the stable and unstable manifolds always
being oriented along the $p$ and $q$ directions, in accordance with
Eq.~\ref{eqmo}. Several system parameters can be easily
varied to collect wavefunction
statistics while keeping fixed the linearized dynamics near a given
periodic orbit. Semiclassical spectra and eigenstates
can be efficiently computed, allowing for comparison between these and
the exact quantum stationary properties~\cite{scbaker}.

In Fig.~\ref{fignumspec}a appears a typical local line
spectrum $S(E)$ (Eq.~\ref{sq0p0}) for a wavepacket centered on a periodic orbit
with exponent $\lambda=0.79$. A total of $223$ states exist in the energy
interval $[-\pi,\pi]$; a portion of this spectrum containing the peak
of the linearized envelope $S_{\rm lin}(E)$ (dotted line) is shown in the
figure. Two intermediate envelopes, defined by cutting off the sum in
Eq.~\ref{spectrum} at $T=30$ (a time long compared to
the orbit period but short compared to the Heisenberg time)
also appear in the figure. The solid curve is obtained from the exact quantum
calculation until $|T|<30$, while the dashed curve comes from the
corresponding semiclassical calculation. We see that the large wavefunction
intensities do appear in the energy region predicted by the linear envelope, 
and further refinement of the spectrum is obtained from intermediate-time
dynamics.

In Fig.~\ref{fignumspec}b we again zoom in on the energy region in which
the linearized envelope (now shown as a solid line) is peaked. The
full spectrum $S(E)$ is still marked by the vertical solid lines, while
a semiclassical line spectrum appears as dashed lines. Though the quantum
and semiclassical spectra differ in their detailed properties, they are both
observed to fluctuate around the same linear envelope $S_{\rm lin}$.
In Fig.~\ref{fignumspec}c we separate out these random fluctuations in the
quantum spectrum from the short-time constraint $S_{\rm lin}$, by
dividing through all of the line heights by $S_{\rm lin}$. We see that the
rescaled line heights display energy-independent oscillations, equally
strong near the peak and valley of the envelope. 

In ~\cite{nlscar} the rescaled line heights of Fig.~\ref{fignumspec}c
are collected for an ensemble of systems, and compared to the Porter-Thomas
prediction of Gaussian random fluctuations. Because we will be focusing on 
wavefunction intensities in the following section, we omit this data here,
but will point out that the IPR for the spectrum of Fig.~\ref{fignumspec}
is $7.93$, which agrees well with the scar theory prediction 
$7.94$ for this value of $\lambda$ when the random fluctuation
factor $F=2$ is included (see Eq.~\ref{scaripr}). The IPR of the
semiclassically evaluated spectrum for this system is $8.46$.

In Fig.~\ref{figiprvslin} we plot the numerically obtained values of the IPR
for a large number of systems, with the test wavepacket placed
on orbits of varying exponent $\lambda$ ($0.28 < \lambda <1.94$). The
actual value of the IPR is plotted versus the linear IPR prediction, 
$<S_{\rm lin}^2>$, i.e. the amount of fluctuation in the linear
envelope.
The latter is obtained
as in Eq.~\ref{scaripr} by summing linearized return probabilities
for a wavepacket centered on an orbit of exponent $\lambda$. Three
sets of data points are shown in the figure (see~\cite{nlscar}),
and a line of slope $F=2$
is plotted for comparison. The data is consistent with the IPR being given
by a product of the fluctuations in the linearized envelope itself
with Porter-Thomas fluctuations under the envelope. The spread of data
points in the figure is also consistent with $O(1/\sqrt{N_{\rm eff}})$ 
fluctuations in the IPR from system to system, where $N_{\rm eff}$
is the effective number of states under the linearized envelope~\cite{nlscar}.

In Fig.~\ref{fig2ptcorrel} we plot the 2-point correlation function
$<S(E+\Delta E)S(E)>/<S(E)>^2$
of the spectrum shown in Fig.~\ref{fignumspec}a (plusses). In the same figure
is plotted (with diamonds) the 2-point correlation function
of the rescaled spectrum of Fig.~\ref{fig2ptcorrel}c. Again, we see that
all of the non-random structure in the spectrum arises from the presence
of the linearized envelope $S_{\rm lin}$, and that the rescaled spectral
lines are completely uncorrelated.

\section{Wavefunction intensity statistics}
\label{wis}

As an example of the calculations which can be easily performed within the
framework of scar theory, we discuss the tail of the wavefunction
intensity distribution, and its deviations from the exponential tail of
RMT\cite{sscar}.
This quantity has been analyzed previously for disordered
systems, where a log-normal tail is predicted~\cite{diffwis}
and observed~\cite{lupusax}.
That analysis applies to scatterers that are small compared to
the Fermi wavelength of the system. In the semiclassical limit,
where the scatterer size is kept fixed while the wavelength is taken to zero,
one obtains very different behavior, as we shall see below. This suggests
that although RMT is known to be a good zeroth-order approximation
for both quantum chaotic and diffusive behavior, the corrections to RMT
may be qualitatively very different in the two kinds of systems.

As we saw above in Section~\ref{lnls}, individual spectral lines in the
local density of states $S(E)$ show, in the 
semiclassical limit, chi-squared (Porter-Thomas) fluctuations
around the linear envelope $S_{\rm lin}$. For complex
eigenstates (in the absence of time-reversal symmetry)
the chi-squared distribution has two degrees of freedom, and
without scarring effects
the probability of having a spectral line height greater
than $x$ is given by $P(x)=\exp(-x)$. Here the intensity
$x$ is normalized to have a mean
value of unity, i.e. $x_n=N |\langle n|a^{q_0,p_0}\rangle|^2$, where 
$|a^{q_0,p_0}\rangle$ is the Gaussian wavepacket and $N$
is the total
number of states ($N=1/h$ for a classical phase space area of unity).
Including scarring effects for an orbit of instability exponent $\lambda$, 
this distribution is modified to
\begin{equation}
P(q_0,p_0,\sigma,\lambda,E,x)=\exp{(-x/S_{\rm lin}(q_0,p_0,\sigma,\lambda,E))}
\end{equation}
for a eigenstate with energy $E$ overlapping with a Gaussian
centered at $(q_0,p_0)$ with width $\sigma$.
Thus, the average intensity becomes
$S_{\rm lin}(q_0,p_0,\sigma,\lambda,E)$, and the entire distribution
is stretched out by this factor.
Now we perform an energy average,
remembering that for a map the quasi-energy $E$ is defined to lie between
$0$ and $2\pi$ only. We notice that the tail of
the intensity distribution will be dominated by the peak of
the spectral envelope at the EBK quantization energy $E=\phi$.
Without loss of generality we may set $\phi=0$, and perform the
energy integral using the saddle point approximation around $E=0$,
where $S_{\rm lin}$ is maximized. We then
obtain
\begin{eqnarray}
P(q_0,p_0,\sigma,\lambda,x) & = & {1\over 2 \pi}
\int dE P(q_0,p_0,\sigma,\lambda,E,x) \nonumber \\
& \approx & {1 \over \sqrt{2 \pi}}
{\exp(-x/S_{\rm lin}(q,p,\sigma,\lambda,E=0))
\over \sqrt{{-x \over S_{\rm lin}(0)^2}
{\partial^2 S_{\rm lin}\over \partial E^2}}} \,,
\end{eqnarray}
valid in the limit of large $x$.
For small $\lambda$, the sum over time steps can be replaced by
an integral, and we have at $q_0=p_0=0$
\begin{equation}
S_{\rm lin}(E)=\int dt {e^{-i Et} \over \sqrt{\cosh{\lambda t}}} \,.
\end{equation}
Now by dimensional analysis, $S_{\rm lin}(0)=Q/\lambda$
and ${\partial^2 S_{\rm lin}\over \partial E^2}(0)=-W/\lambda^3$,
where $Q$ and $W$ are numerical constants. We thus obtain
the $\hbar-${\it independent} tail of the intensity distribution
for a wavepacket centered on a periodic orbit,
\begin{equation}
\label{res1}
P(q_0=0,p_0=0,\sigma,\lambda,x) =
{1 \over \sqrt{2 \pi}} {Q \over \sqrt W} \lambda (x \lambda)^{-1/2}
 e^{-x\lambda/Q} \,.
\end{equation}
This is the total probability of having a wavefunction intensity
enhanced by a factor $x$ over the average intensity, on a periodic orbit
of instability exponent $\lambda$.
The result is independent of $\sigma$, and thus of the aspect ratio
of the test state in phase space.
Notice that the exponential tail
has been effectively stretched by a factor of $Q/\lambda$, corresponding
to the height of the peak of the linear envelope at small
$\lambda$. There is also a linear suppression factor of $\lambda$,
corresponding to the energy width of the peak in $S_{\rm lin}(E)$, and
indicating that only a fraction scaling as $\lambda$ of all the eigenstates
are effectively scarred. See the upper curve in Fig.~\ref{figwis1} for a
comparison of this result with numerical data.

The region of validity of Eq.~\ref{res1} is
\begin{equation}
1 \ll \lambda^{-1} \ll x \ll N \,.
\end{equation}
The
first inequality ensures that many iterations of the periodic orbit
contribute (so the sum over iterations can be replaced by
an integral) and the scarring is strong. In fact, however, because of
the large value of the numerical constant
$Q \approx 5.24$, the formula works well
even for exponents as large as $\log 2$, as will be seen in the
numerical study below. The second inequality  says that we are in the
tail of the distribution and the events are all coming from the
peak of the linear envelope $S_{\rm lin}$. The third inequality is a unitarity
constraint: obviously our assumption of random fluctuations breaks
down for intensities of order $N$, when the entire wavefunction
would be concentrated in a phase space area of order $\hbar$.

A similar analysis can be performed integrating over the phase
space variables $q_0$ and $p_0$, to obtain the full distribution 
of wavefunction intensities everywhere in phase space.
The exponential $\exp{(-x/S_{\rm lin}(q_0,p_0,\ldots))}$
must now be expanded to second order in $q_0,p_0$
around the maximum $q_0=p_0=0$.
Then upon integration by the saddle point method we obtain
a determinant
factor~\cite{sscar} of ${\pi \hbar\over (x\lambda)} {Q^2\over Z}$, again
independent of
$\sigma$. $Z$ is another numerical constant, related to the falloff of 
$S_{\rm lin}$ away from the fixed point $q_0=p_0=0$. The factor
of $\hbar$ in front results from the fact that the dominant contribution
to the tail comes from a region scaling as $\hbar$ surrounding the
periodic orbit (compare Eq.~\ref{alinoff} and the discussion following).
Combining this determinant prefactor
with the expression in Eq.~\ref{res1},
we obtain
the tail of the distribution
of overlap intensities after energy and phase space averaging,
\begin{equation}
\label{res2}
P(\lambda,x)=
\sqrt{\pi \over 2} {Q^3 \over Z \sqrt W} \hbar \lambda
(x\lambda)^{-3/2}\exp{(-x\lambda/Q)} \,.
\end{equation}
This result is valid in the regime
\begin{equation}
\max(\log N,\lambda^{-1}) \ll x \ll N \,.
\end{equation}
Here $\log N$ is the value of intesnity $x$ near which the RMT exponential
decay law $\exp(-x)$ reaches values of order $\hbar=1/2\pi N$.
Near this value of $x$, a crossover occurs between the head
of the distribution, which is dominated by non-scarred region
of phase space and approaches the Porter-Thomas (RMT)  prediction,
and the tail, dominated by scarring,  given by the expression above.
The  expression Eq.~\ref{res2} holds also for an ensemble of
systems, all having one orbit with instability $\lambda$. In principle,
we should of course do a sum over all periodic orbits in a given
system, however the tail will clearly
always be dominated by the orbit with smallest $\lambda$. Numerical data
confirming Eq.~\ref{res2} appears in the lower curve
of Fig.~\ref{figwis1}.

Finally, we consider an ensemble of systems
where the value of the smallest
exponent $\lambda$ varies from realization to realization,
with distribution
${\cal P}(\lambda)=C \lambda^\alpha$ for small $\lambda$.
An example would be a billiard system with many movable disks, each of
diameter much larger than the wavelength.
Then using Eq.~\ref{res2} and integrating over  $\lambda$ we obtain
\begin{equation}
\label{res3}
P(x)=C \sqrt{\pi \over 2} {Q^3 \over Z \sqrt W}
Q^{\alpha+1/2}\Gamma(\alpha+1/2)
\hbar x^{-(2+\alpha)} \,.
\end{equation}
Note that this is an uncontrolled approximation because
we have integrated over $\lambda$ after having assumed
$x \lambda$ to be large. However, if we had included higher-order
corrections in $(x \lambda)^{-1}$ in Eq.~\ref{res2}, the scaling of
$P(x)$ would remain unchanged, i.e.
\begin{equation}
P(x)= C f(\alpha) \hbar x^{-(2+\alpha)} \,,
\end{equation}
with the dimensionless function $f(\alpha)$ somewhat different
from that given in Eq.~\ref{res3}. The important point is that the
tail displays power-law behavior in the intensity $x$, a strong
deviation from the exponential prediction of RMT. As with Eq.~\ref{res2},
this asymptotic form is valid for values of $x$ large compared to
$\log N$ and small compared
to $N$. For small $x$ we again
expect a crossover to the Porter-Thomas form $\exp(-x)$. For large $x$
we expect a downward correction away from  the $x^{-(2+\alpha)}$ form,
with a breakdown of the approximation occurring at some
fraction of $N$, depending on $\alpha$. 

We now present numerical evidence for the results of this section, using the
same ensemble of systems as in the numerical tests of the previous section.
The data shown here was originally obtained in \cite{sscar}.
In Fig.~\ref{figwis1} we fix the instability exponent  $\lambda$ 
of the least unstable periodic orbit at
$\log 2$, and obtain a cumulative wavefunction intensity distribution for
a wavepacket centered on the orbit (upper thick curve), and a combined
distribution for wavepackets randomly located over the entire phase space
(lower thick curve).
The tails of the distributions compare well with the scar theory predictions
of Eq.~\ref{res1} (dashed curve) and Eq.~\ref{res2} (solid curve). Notice
that the generic wavepacket wavefunction intensity distribution (lower
thick curve) rolls over to the RMT prediction (dotted line) at small
values of the intensity. The log-time in this system, near which value
of $x$ the crossover
is expected to occur is $T_{\rm log} \approx 7$.

In Fig.~\ref{figwis2} the same distribution is shown after 
ensemble averaging over systems with
classical orbits of different instability exponents. We see that at small
values of $x$ the data agrees well with the RMT prediction (dashed line), while
the tail follows the power-law behavior of Eq.~\ref{res3} (solid line).

\section{Improved (universal) measures of scarring}
\label{ims}

We have noted previously an ambiguity in the choice of Gaussian wavepacket
to serve as a test state for measuring scarring effects. Even after requiring
that the wavepacket be optimally oriented relative to the invariant manifold
directions, a one-parameter family of possible Gaussians remains,
parametrized by the aspect ratio $\sigma^2$ (See Eq.~\ref{wavepkt}).
Together, these fill a hyperbolic region $|qp| \le  O(\hbar)$
encompassing the stable and
unstable axes: the scarring phenomenon is present in this entire 
region and not just at the periodic orbit $q=p=0$ itself. [Compare this
with the hyperbolic fringes in the Wigner spectral function obtained
by Berry~\cite{berryscar}]. This fact suggests that the Husimi-space
ambiguity may be resolved, and a better measure of scarring obtained, 
by including information about wavefunction behavior in this entire
hyperbolic region instead of choosing one Gaussian of arbitrary
width $\sigma$.

A similar ambiguity arises when we consider a higher-period orbit of a map,
or a periodic orbit of a continuous-time system. There, the test wavepacket
is chosen to be centered at one of the many possible periodic points;
again a less arbitrary measure of scarring should be obtainable by
considering the wavefunction along the entire orbit.

The obvious way to resolve these ambiguities is by averaging scarring
intensity over wavepackets of different aspect ratios in phase space,
or over wavepackets
centered at  different periodic points, or both. Thus, we replace the
single wavepacket $|a\rangle$ by a density operator $\rho$, and
define the scarring strength  for a given eigenstate $|n\rangle$  as
\begin{equation}
I_{n \rho} \equiv \langle n | \rho | n \rangle \,.
\end{equation}
Then we can construct a wavepacket-averaged local density of states analogous
to Eq.~\ref{sq0p0}:
\begin{equation}
S_\rho(E)=\sum_n I_{n \rho} \delta(E-E_n) \,,
\end{equation}
and a corresponding inverse participation ratio
\begin{equation}
{\rm IPR}_\rho = N \sum_n I_{n \rho}^2 \,.
\end{equation}

To obtain averaging over the wavepacket aspect ratio at a given periodic
point, we write
\begin{equation}
\label{rhodef}
\rho = {\cal N} \int dt \, e^{-t^2/T_0^2} \,
|a_{\sigma e^{\lambda t}} \rangle \langle
a_{\sigma e^{\lambda t}} | \,.
\end{equation}
The exponent $\lambda$ can be chosen to be the instability exponent of the
unstable orbit (Eq.~\ref{eqmo}), but this choice is in fact arbitrary,
and $\lambda$ can be reabsorbed into the overall normalization factor $\cal N$
and the time cutoff $T_0$. In the absence of the cutoff $T_0$, the hyperbolic
test state would be completely scale invariant, giving equal weight
to Gaussians of all aspect ratios, from tall and thin to short and wide.
The cutoff is, however, necessary because the linearized dynamics of
Eq.~\ref{eqmo} is in fact only valid in a finite
classical region around the periodic
orbit (and also it eliminates possible normalization difficulties).
To get an optimal measure of scarring, we choose the initial width 
$\sigma$ such that the wavepacket has the same aspect ratio in phase space
as the
linearizable region surrounding the orbit in which the equations
of motion are well described by Eq.~\ref{eqmo}. Then the initial
wavepacket $|a_\sigma\rangle$
can expand by the same stretching factor along either of the two invariant
directions before hitting the ``edge" of the linearizable region.
One easily sees that if $A$ is the phase-space area of the linearizable region,
we want to choose
\begin{equation}
\label{t0def}
T_0= {c \over 2 \lambda} \log {A \over 4 \hbar} \,.
\end{equation}
$c$ is an arbitrary constant of order unity: it determines
the exact suppression factor in Eq.~\ref{rhodef} for a Gaussian just touching
the boundary of the linearizable region. In the semiclassical limit, of course,
$A \gg \hbar$, and the precise size $A$ of the region needs to be 
determined only to within a multiplicative factor; it may indeed in this limit
be taken
as equal to the volume of the entire accessible phase space. Notice
that up to a constant, $T_0$ has the same form as the log-time
$T_{\rm log}$ defined in Section~\ref{lnls}. The Gaussian
form of Eq.~\ref{rhodef} in the time variable $t$ is somewhat arbitrary,
and also not important in the $\hbar \to 0$ limit. See \cite{is}
for a fuller discussion.

To average over the position of the periodic point, we similarly may define
a phase-space ``tube"
\begin{equation}
\label{tube}
\rho= {\cal N} \int dz  \, |a_{z,\sigma,\sigma_z} \rangle
\langle a_{z,\sigma,\sigma_z}| \,,
\end{equation}
where the $z$ coordinate parametrizes the periodic orbit in phase space,
and at each periodic point the wavepacket is chosen to have width $\sigma_z$
along the direction of the orbit and width $\sigma$ in the unstable
direction at that point on the orbit. We recall that the Husimi function
is nothing other than a Gaussian smoothing of the Wigner spectral
function; thus averaging the Husimi function around the periodic orbit
coincides with the phase-space tube averaging of Fishman, Agam,
et al.~\cite{fishman}. In \cite{fishman}, the Gaussian width $\sigma$
is generally
taken to be fixed as one goes around the tube; however that is a
coordinate-dependent condition, and in general one may take $\sigma$ to
be an arbitrary function of $z$. 

More generally one can, of course, combine  the averaging procedures
of Eqs.~\ref{rhodef},\ref{tube}, to obtain a fully-universal measure
of scarring. In the strong-scarring limit $\lambda \ll 1$ one may also
without loss of generality begin with one wavepacket only and allow
the dynamics itself to generate the density matrix $\rho$:
\begin{equation}
\label{rhodyn}
\rho_{\rm dyn} = \sum_{t=-\infty}^{+\infty} \,
e^{-t^2/T_P^2 T_0^2} \, |a_{\rm lin}(t)\rangle \langle
a_{\rm lin}(t)| \,.
\end{equation}
Here $|a_{\rm lin}(t)\rangle$ is the original Gaussian $|a\rangle$ evolved in
accordance with the {\it linearized} dynamics: for example, if $t$
is an integer multiple of the period $T_P$, then
$|a_{\rm lin}(t)\rangle$ is centered at the same periodic point as
$|a\rangle$, but with width $\sigma \to \sigma e^{\lambda t/T_P}$.
$T_0$
is defined as before (Eq.~\ref{t0def}), using the full instability
exponent $\lambda$ for one iteration of the {\it entire} primitive orbit.
$\lambda/P$ is the exponent {\it per time step}; hence the factor of $T_P^2$
in Eq.~\ref{rhodyn}. 

To understand what gains may be achieved by the averaging process
described here, we need to consider correlations in the local
density of states
between different wavepackets placed on the same periodic orbit. Indeed,
these correlations will also be very important when we discuss
conductances and  phase space transport in Section~\ref{open}.
To begin, consider two wavepackets $|a\rangle$ and $|b\rangle$
and their overlap intensities $I_{na}$, $I_{nb}$ with the
eigenstates $|a\rangle$
of the system. We may define a
long-time averaged transport probability $P_{ab}$ \cite{wqe} as
\begin{equation}
P_{ab} = \lim_{T \to \infty} {1 \over T} \sum_{t=0}^{T-1}
|\langle a | U^t |b \rangle |^2 \,.
\end{equation}
For a nondegenerate spectrum we easily see
\begin{equation}
P_{ab} = \sum_n \, |\langle a | n\rangle|^2 \, |\langle b | n\rangle|^2
= \sum_n I_{na} I_{nb} \,.
\end{equation}
In particular, the IPR corresponds to the special case
\begin{equation}
{\rm IPR}_a = N P_{aa}  = N \sum_n I_{na}^2 \,.
\end{equation}
The $P_{ab}$ can be thought of as the covariance matrix of the local densities
of states for different wavepackets, with $P_{aa}$ being the variances or
diagonal matrix
elements; the correlation between two local densities of states is then
given by
\begin{equation}
\label{cabdef}
C_{ab} = {P_{ab} \over \sqrt{ P_{aa} P_{bb} }} \,.
\end{equation}

Consider first the case where $|a\rangle$ and $|b\rangle$ are exact
time iterates of one another: $|b\rangle=|a(t)\rangle$ for some time $t$.
Then the two local densities of states are identical,
\begin{equation}
P_{ab}=P_{aa}=P_{bb}={F\over N} g(\lambda)
\end{equation}
and the correlation is unity. $g(\lambda) \approx \pi/\lambda$ for small
$\lambda$ is the scarring IPR enhancement factor (Eq.~\ref{scaripr}).
On the other hand, if we consider two wavepackets that have the same
short-time behavior  but are otherwise unrelated (e.g. if
$|a\rangle$ and $|b\rangle$ are located on different periodic orbits of the
same length and instability exponent $\lambda$), then the two have
the same linearized spectral envelope but independent fluctuations under
that envelope. Then
\begin{eqnarray}
P_{aa}&=& P_{bb} ={ F \over N} g(\lambda) \nonumber \\
P_{ab}&=& { 1\over N} g(\lambda) \nonumber \\
C_{ab} &=& {1 \over F} \,.
\end{eqnarray}
The correlation in this case is of order unity but still less than one.

In the intermediate case, we may have two wavepackets one of which {\it partly}
overlaps the other under short time evolution. Then the random spectral
fluctuations under the two identical envelopes are partially correlated,
and we get a correlation $1/F < C_{ab} < 1$ \cite{is}. However, 
for small $\lambda$, {\it any} two wavepackets optimally oriented along the
same periodic orbit (whether centered at the same periodic point or not)
will overlap each other with a very large amplitude under time evolution.
The reason is that the wavepacket stretches very little each orbit iteration,
and there will always be some $n$ for which $\sigma e^{n \lambda}$
is very close to any chosen width $\sigma'$. Expressions for
determining the LDOS correlation $C_{ab}$ given $\lambda$
and the relative widths of the two wavepackets are given in 
\cite{is}, where the predictions are also numerically tested
[Fig.~\ref{figiscab}]. It is found that
a scarring enhancement factor $g(\lambda)$ of $2$ (corresponding to
instability exponent $\lambda \approx \log 5$), is already associated
with a {\it minimum} correlation of $0.99$ between the least
correlated wavepackets on such an orbit. Orbits that have smaller
$\lambda$ and are thus more scarred have correlations $C_{ab}$ even
closer to unity.

In Fig.~\ref{figiscab} we plot the predicted and numerically observed
correlations $C_{ab}$ between local densities of states for two wavepackets
centered on the same periodic orbit. The parameter $z$ is defined such
that wavepacket $|a\rangle$, under linear evolution, evolves to be centered
at the same point as $|b\rangle$, but with width $e^{\lambda(n+z)}$ times
that of $|b\rangle$, where $n$ is an arbitrary integer. For integer $z$
the two local densities of states are the same, and the correlation is
unity. The smallest correlation is obtained for half-integer $z$. The
upper and lower curves in the figure correspond to instability exponents
$\lambda=\log 5$ and $\lambda=\log 10$, respectively. The data is obtained
by averaging over an ensemble of kicked-baker maps~\cite{is}, and the
theoretical curves are obtained from \cite{is}. We notice that even for these
very moderate instability exponents, the minimum value of $C_{ab}$ for
the most uncorrelated wavepackets is already very close to one. The
correlations get even closer to one as $\lambda$ decreases.

Thus in the strong scarring limit $\lambda \ll 1$, there is in fact no
ambiguity in choosing either the central point $z$ or the aspect
ratio $\sigma^2$ of the test Gaussian on a periodic orbit. This also means
that the density matrix averaging outlined above {\it does not}
produce a scarring measure different from that given by
any one arbitrary wavepacket along the orbit.

Yet it is possible to use our knowledge of the entire orbit and the
linearized classical dynamics surrounding it to build an improved
scarring test state. Beginning once again with a fixed point
of a discrete-time map, we construct the {\it coherent}
sum
\begin{equation}
\label{cohdefth}
\Psi={\cal N} \int dt \, e^{-t^2/T_0^2} e^{i \theta t}\,
|a_{\sigma e^{\lambda t}} \rangle
\end{equation}
(compare with
Eq.~\ref{rhodef}). Again, $T_0$ is a cutoff time (Eq.~\ref{t0def}) 
for the linearized
dynamics, and ${\cal N}$ is a normalization constant. $\theta$ is an
arbitrary phase accumulated per time step. We notice that if the
linear equations of motion (Eq.~\ref{eqmo}) were exact, we could remove
the cutoff $T_0$, and the state $\Psi$ would be an exact eigenstate of the
dynamics, with energy $\phi+\theta$. The extent to which $\Psi$ fails
to be an exact eigenstate is determined by the nonlinear nature
of the dynamics beyond the log-time $T_0 \sim |\log \hbar|/\lambda$,
whereas the failure
of a single wavepacket $|a\rangle$ to be an eigenstate results already
from the initial decay at the time scale $1/\lambda$. Just as scarring
strengths and IPR's for a single wavepacket scale as $1/\lambda$ because
the autocorrelation function $A_{\rm lin}$ remains of order unity
for the first $O(1/\lambda$) iterations of the orbit, so here
the autocorrelation function should decay on a time scale of
order $T_0 \sim |\log \hbar|/\lambda$ in the $\hbar \to 0$ limit, and scarring
strengths should scale likewise. 

Indeed, in the strong-scarring semiclassical regime ($\lambda \to 0$
and $\hbar \to 0$) we obtain~\cite{is} the short-time autocorrelation
function for $\Psi$
\begin{equation}
\label{alinth2}
A_{\rm lin}^\Psi (t)=e^{-i (\phi+\theta) t} e^{-t^2/T_0^2} \,.
\end{equation} 
Fourier transforming, we then have the spectral envelope
\begin{equation}
\label{sth2}
S_{\rm lin}^\Psi(E)=\sqrt{2 \pi} T_0 e^{-(E-\phi-\theta)^2T_0^2/2} \,,
\end{equation}
centered at energy $E=\phi+\theta$, with width scaling
as $\lambda/|\log \hbar|$, and height and IPR [Fig.~\ref{figisipr}] scaling as 
$|\log \hbar|/\lambda$. In particular, the inverse particpiation ratio 
has the following asymptotic form for small $\lambda$ and small $\hbar$:
\begin{equation}
\label{iprpsi2}
{\rm IPR}_\Psi = F \sqrt \pi T_0\,.
\end{equation}
Note that the LDOS of a single
wavepacket is always peaked at the ``quantization energy" $E=\phi$,
whereas by taking coherent superpositions we may construct an
``off-resonance" state with a LDOS peaked away from this energy
[see Fig.~\ref{figisth}].
For $\theta=0$, the state $\Psi$ lives on the fixed point itself and on
the invariant manifolds within the linearizable region around the orbit;
for non-zero $\theta$ it lives on hyperbolic regions asymptotic to the two
manifolds [Fig.~\ref{figishus}]:
\begin{equation}
\label{hypstruct}
{qp \over \hbar} \sim {\theta /\lambda}
\end{equation}
These hyperbolic structures living near the periodic orbits
but not on the orbits themselves 
have been studied previously by Nonnenmacher and Voros~\cite{voros},
and observed before that by Saraceno~\cite{sarbaker} and
by O'Connor and Tomsovic~\cite{octbaker} in the context of the baker
map.
The association between energy shifts and hyperbolic structures
surrounding the orbit is also evident in the phase of the dynamics
in Eq.~\ref{alinoff} for wavepackets centered off of the periodic orbit.
We see there that at times short compared to $\lambda^{-1}$, the phase
accumulated per time step is shifted by the amount $q_0p_0 \lambda/\hbar$
(where $(q_0,p_0)$ is the center of the wavepacket), and the peak of
the spectral envelope in quasienergy space must therefore also be shifted
by an amount of this order.

The scarring measure given by Eq.~\ref{cohdefth} may truly be said to
be universal and optimal, as it eliminates the ambiguity inherent
in the Husimi space measure, and
fully incorporates all knowledge not only about the
location of the periodic orbit itself but also about the linearized
dynamics around this orbit. 
We note here that the expressions Eqs.~\ref{alinth2},\ref{sth2} are valid 
in the $\lambda \to 0$ limit, where scarring is strong. However, this improved
measure of scarring can also be used for moderate $\lambda$, where 
the expressions (Eq.~\ref{alinth2},\ref{sth2}) need to be modified.
Although the dependence of the spectral envelope on $\lambda$ is more
complicated away from the $\lambda \ll 1$ regime, the spectral envelope
height and the IPR still scale as $\log\hbar$ in the semiclassical limit,
for arbitrary $\lambda$. Thus, even an unstable orbit with moderate
or large instability exponent, which will be rather weakly scarred
using the usual Gaussian wavepacket measure, will always be strongly
scarred using the ``universal scarmometer" $\Psi$, in the $\hbar \to 0$
limit. See \cite{is} for a more thorough discussion of this point, as
well as of finite $\hbar$ corrections to theoretical predictions of
scar strength.

We now briefly discuss the extension of the above methods to longer
periodic orbits of maps and to continuous time. Here, even greater
enhancement of scarring intensity is obtained because the universal
test state $\Psi$ is naturally defined to live along the {\it entire}
orbit. For a map of period
$T_P$, we recall that there are $T_P$ optimal energies
\begin{equation}
E_k= {\phi + 2\pi k \over T_P} \; \; , \; \; k=0 \ldots T_P-1 \,.
\end{equation}
that ``quantize" the orbit in the EBK sense. The spectral envelope
of a single Gaussian wavepacket will be peaked at all these $T_P$
energies (see the discussion following Eq.~\ref{shorthp}). We now
choose one such quantization energy and construct an optimal
scarring test state $\Psi$ centered at energy $E_k+\theta/T_P$:
\begin{eqnarray}
\label{hiperpsi}
\Psi & = &{\cal N} \sum_p \int dt e^{-t^2/T_P^2 T_0^2} \nonumber \\
& \times & e^{i (E_k + \theta/T_P)p
+i\theta t/T_P -i \phi_p }
|a_{x_p,\sigma f_p e^{\lambda t/T_P}} \rangle \,.
\end{eqnarray}
Here $f_p$ is a stretching factor, and $\phi_p$ is a phase, both
defined by
\begin{equation}
U_{\rm lin}^p |a_{x_0,\sigma} \rangle = e^{-i \phi_p}|a_{x_p,\sigma f_p
e^{\lambda p/T_P}}\rangle  \,.
\end{equation}
$f_p$ and $\phi_p$ take into account the fact that stretching and
phase
accumulation along the orbit may both be non-uniform; of course,
$f_{T_P}=1$ and $\phi_{T_P}=\phi$.
The correct phases
$\phi_p$ are particularly crucial for getting constructive
interference between different periodic points.
Once again, $\theta$ is an arbitrary off-resonance parameter. For
$\theta=0$, the test state lives on all $T_P$ periodic points and
their invariant manifolds; for $\theta =O(\lambda)$, the test state
occupies hyperbolic regions $qp \sim \hbar$ near these
same periodic points.

The short-time autocorrelation function and the spectral envelope
take the same form as in the $T_P=1$ fixed point case, making the 
substitutions
\begin{equation}
T_0 \to T_0 T_P \;,\;\lambda \to \lambda /T_P \;,\;
\theta \to \theta /T_P  \;,\; \phi \to E_k
\end{equation}
throughout. Notice in particular that the relevant instability parameter
is $\lambda/T_P$, the exponent {\it per time step}, and not the exponent
$\lambda$ per iteration of the entire orbit which governs the strength
of the single-wavepacket scarring measure. In particular, the peak height
in the LDOS $S_{\rm lin}^\Psi(E)$ and the corresponding IPR both scale
as $(\lambda/T_P)^{-1}|\log\hbar|$ for small $\lambda/T_P$.
Thus, as long as the Lyapunov
exponent $\lambda/T_P$ remains small, the effects of longer orbits (up
to the mixing time) can remain strong using the $\Psi$ test-state measure,
whereas the single-wavepacket measure will fail to detect a strong
signal from these longer orbits. For $T_P>1$, a dramatic increase
in scarring strength is observed even for moderate $\hbar$, because
of this additional $T_P$ factor.

For small $\lambda$, the optimal test state $\Psi$ can also
be generated dynamically:
\begin{equation}
\label{hiperdyn}
\Psi_{\rm dyn} \sim \sum_t \, e^{-t^2/T_P^2 T_0^2} \,
e^{i(E_k+\theta/T_P)t} |a_{\rm lin}(t) \rangle\,
\end{equation}
(compare Eq.~\ref{rhodyn}). Here $|a \rangle$ is a wavepacket of width $\sigma$
centered at {\it any} point along the periodic orbit.

The expressions Eqs.~\ref{hiperpsi},\ref{hiperdyn}
also generalize in an obvious
way to continuous time\cite{is}. Again, one chooses any
energy $E_k$ which quantizes the action in units of $\hbar$, and
an arbitrary energy offset $\hbar \theta/T_P$ if one wishes
to measure off-resonance scars. The short-time autocorrelation
function again decays on the time scale
$T_0 T_P \sim T_P \lambda^{-1} \log {A /\hbar}$. In the energy
domain it produces a {\it single} peak of height scaling as
$(T_P \log {A /\hbar})/\lambda$ compared to the background spectral envelope
(of width $\hbar/T_{\rm free}$, see discussion in Section~\ref{lnls})
which enforces energy conservation. This is to be compared with
the ordinary scarring envelope which has peaks at {\it all} of the
quantization energies with height scaling simply as $1/\lambda$. Once again,
we find that longer orbits can be detected as easily as short ones using
the improved scarring
measure\footnote{Of course, given a value of $\hbar$ only orbits of period less
than the mixing time for that $\hbar$ can be profitably studied; our approach
is based on the assumption $T_P < T_{\rm log}$, so that linearized dynamics
remains valid over several iterations of the orbit.}
and an additional logarithmic scarring enhancement factor
is obtained in the semiclassical limit.

We mention here the closely related work of de Polavieja, Borondo, and
Benito on improved measures of scarring
in Hamiltonian systems~\cite{borondo} and of
Simonotti, Vergini, and Saraceno~\cite{saraceno} on the surface of section
in a billiard system. In both cases, a coherent superposition of wavepackets
is used to construct a better test state for measuring scarring effects.
However, both these groups consider only one (arbitrarily shaped)
wavepacket at each periodic
point of the orbit; thus they obtain the enhancement factor
$T_P$ but not the extra $|\log\hbar|$ factor that comes from using the
linearized dynamics to fill the entire hyperbolic region
surrounding the orbit. See \cite{is} for a fuller discussion of this and other
related issues.

Numerical data in this section is obtained from an ensemble of kicked
baker maps; see \cite{is} for details. In Fig.~\ref{figisipr}, we show
the IPR plotted as a function of the cutoff constant $c$ (see Eq.~\ref{t0def}),
for various values of the system size $N$ ($N=1/h$ is the number of
Planck-sized cells in the classical phase space). Here the instability
exponent for the period-one orbit is $\lambda=|\log 0.18|$.
The upper dashed curve is the $N \to \infty$ theoretical prediction, which
converges to the asymptotic form of Eq.~\ref{iprpsi2} (lower dashed line)
for large $T_0$ (i.e. for exponentially small $\hbar$, keeping
fixed the cutoff parameter $c$). The
linearized theory is expected to start breaking down for $c \ge 1$ (the
rightmost
six points on each data curve).

In Fig.~\ref{figisth} are shown the smoothed (numerically obtained)
local densities of states for off-resonance test states $\Psi$, for several
off-resonance angles $\theta$. The tallest peak corresponds to the
on-resonance test state $\theta=0$; in the $\hbar \to 0$ limit the peaks
would all be the same height and shape. The system size here is fixed at
$N=800$, the cutoff parameter is $c=0.6$, and once again $\lambda=|\log 0.18|$.
In Fig.~\ref{figishus} are displayed Husimi plots of the 
universal test state $\Psi$ for (a) $\theta/\lambda=0.8$ and
(b) $\theta/\lambda=2.5$. For $\theta=0$ (not shown), $\Psi$ lives entirely
on the orbit (located here in the center of the figure) and on the
invariant manifolds (vertical and horizontal lines intersecting at the
center of the figure). As $\theta$ becomes non-zero,
$\Psi$ moves off of the manifolds and onto hyperbolic regions asymptotic
to the manifolds (Eq.~\ref{hypstruct}).
Choosing the opposite sign for $\theta$ would produce a test
state occupying the other two quadrants.

Finally, in Fig.~\ref{figistp2} we show a numerically obtained,
smoothed spectrum for a
wavepacket placed on a period-$2$ orbit (double-peaked solid curve),
and for the universal test state $\Psi$ constructed on the same orbit
(tall single peak). These agree well with the theoretical predictions
(dashed curves) based on the linearized dynamics near the periodic
orbit, which has a total exponent $\lambda=|\log 0.168|$ over the
two-step period.
One of two possible on-resonance energies has been chosen for the
test state $\Psi$, which is again constructed using cutoff constant
$c=0.6$. The enhancement here is more dramatic than for the period-one
orbits, due to the fact that
universal scarring strength depends only on the exponent per unit time
along the orbit, not on the orbit length itself.

\section{Open systems: scars and antiscars}
\label{open}

We now proceed to consider the last major topic to be covered in this
review: the effect of unstable periodic orbits on weakly open
systems, where experimentally accessible quantities such as resonance
peaks and conductance curves can be measured. See \cite{open} for
a more detailed exposition of the effect of periodic orbits on
resonances and on the probability to remain
in an open quantum chaotic system. A study of conductance curves
in two-lead chaotic systems is now in progress~\cite{cond}.

Let $H_0$ be a quantum
Hamiltonian for a classically chaotic (closed) system, and consider
adding a small opening to the system, through which probability
can escape. If the opening size corresponds to less than one open
channel, so that quantum resonances will be non-overlapping,
we can write an effective Hamiltonian for the open
system:
\begin{equation}
\label{hdef}
H=H_0 -i{\Gamma \over 2} |a\rangle \langle a| \,.
\end{equation}
Here $|a\rangle$ is a quantum channel associated with the opening, and
$\Gamma$ is the decay rate in that channel. $|a \rangle$ could be a
Gaussian wavepacket at the location of the hole, or a position
or momentum state. The main effect of the opening on
a wavefunction
$|n\rangle$ of the closed system is that it acquires a decay width
proportional to the intensity of the wavefunction at the opening:
\begin{equation}
\Gamma_n = \Gamma |\langle n|a\rangle|^2 \,.
\end{equation}

In RMT, the intensities $x_n \equiv N|\langle n|a\rangle|^2$
follow a chi-squared distribution: $P(x)={1 \over \sqrt{2\pi x}}
\exp{(-x/2)}$ for real overlaps, and $P(x)=\exp{(-x)}$ if the overlaps
$\langle n|a\rangle$ are complex. On the other hand, if the decay
channel $|a\rangle$ happens to be located on a short periodic orbit,
we know that the distribution of decay rates is stretched by the
factor $S_{\rm lin}^a(E)$, this being the smoothed LDOS at $|a\rangle$.
[Compare with the discussion of wavefunction intensity statistics in
Section~\ref{wis}.]
This will have a profound effect on the distribution of resonance
lifetimes in the system. Near the would-be EBK quantization energies
where $S_{\rm lin}>1$, resonance lifetimes are decreased (by a factor
scaling as $\lambda^{-1}$ for small $\lambda$), while far from these energies
resonances tend to be narrow compared with the expectations of RMT. In
order to have a concrete quantity to focus on, let us consider the
ensemble-averaged probability to remain in the system at long
times. This quantity has been investigated previously for
disordered systems~\cite{miller},
and a log-normal long-time tail was found.
We investigate here what deviations from RMT predictions can be obtained
as a result of periodic orbit effects.

First consider the RMT case, with complex eigenstates.
Because mixing between the states of the closed system can be neglected
in the small $\Gamma$ regime, the total probability to remain in the system
is given by a sum over these states:
\begin{eqnarray}
P_{\rm rem}(t) & = & {1 \over N} \sum_{n=0}^{N-1} e^{-{x_n \over N} \Gamma t}
= \int_0^\infty dx P(x) e^{-x \Gamma t/N} \nonumber \\
& = & \int_0^\infty dx e^{-x} e^{-x \Gamma t/N}
= {1 \over 1+ \Gamma t/N} \,.
\label{rmtdec}
\end{eqnarray}
Here $N$ is the total number of states in the system; the classical
decay rate is given by
\begin{equation}
\Gamma_{\rm cl} = \Gamma/N
\end{equation}
because only one channel has the possibility
to decay.
We see that at short times ($t \ll \Gamma_{\rm cl}^{-1}$), the
probability to remain in the system is $P_{\rm rem}(t)
\approx 1-\Gamma_{\rm cl} t$,
as expected, while at long times the power-law asymptotic behavior
\begin{equation}
P_{\rm rem}(t) \approx { 1 \over \Gamma_{\rm cl} t} \,.
\end{equation}
is obtained. The analysis can be extended to $M>1$ open channels,
where the power of the long-time behavior scales with $M$. Notice
that at the classical level we expect an exponential decay
law $P_{\rm rem}(t)=\exp(-\Gamma_{\rm cl} t)$, since the system is chaotic
and the hole small enough that the probability distribution inside the system
never deviates from uniformity (the mixing time being very short compared
to the decay time).

Now take the open channel $|a\rangle$ to be centered on an
unstable periodic orbit of exponent $\lambda$. The intensities
are then distributed in each energy range as a
chi-squared distribution with mean $S_{\rm lin}(E)$:
\begin{equation}
P(x)={1 \over S_{\rm lin}(E)} e^{-x/S_{\rm lin}(E)} \,.
\end{equation}
Then the probability to remain at long times is given by
\begin{eqnarray}
\label{preme}
P_{\rm rem}(t) & = & \int_0^\infty dx P(x) e^{-x \Gamma t/N} =
{1 \over 1 + S_{\rm lin}(E) \Gamma t/N} \nonumber \\ & \to &
{1 \over S_{\rm lin}(E)}  { 1 \over \Gamma_{\rm cl}t}
\end{eqnarray}
if initially only states with energy around $E$ are populated. Three
qualitatively different energy regimes can be distinguished
[Fig.~\ref{figdecslin}].
At the quantization energy $E=\phi$, $S_{\rm lin}(E)$ reaches its peak;
the peak height scales as $\lambda^{-1}$ for small $\lambda$,
and its width scales as $\lambda$ compared to the peak spacing. 
This is in fact the energy region which dominates the tail of the wavefunction
intensity distribution in Section~\ref{wis}. Away from the
peak, for $|E-\phi| \gg \lambda$, the linearized LDOS
falls off exponentially:
\begin{equation}
\label{inters}
S_{\rm lin}(E) \approx {2 \pi \over \lambda} e^{-\pi |E-\phi| /2 \lambda} \,.
\end{equation}
[Notice that for simplicity the above expression and those to
follow are given in quasienergy units, where the
spacing between successive
scarring peaks is fixed at $2\pi$. A factor of $T_P/\hbar$,
where $T_P$ is the orbit period in real time units, needs to be inserted to
convert $E$ to a real energy.]

Within $O(\lambda)$ of the antiscarring energy $E=\phi+\pi$ (this is the
energy equidistant between two successive scarring peaks),
the LDOS rolls over smoothly to attain the minimum value
\begin{equation}
\label{mins}
S_{\rm lin}(E=\phi+\pi) \approx {4 \pi \over \lambda} e^{-\pi^2/2 \lambda}\,.
\end{equation}
The region within $O(\lambda)$ of the antiscarring energy
$E=\phi+\pi$ is thus
responsible for producing the smallest wavefunction intensities, and
the narrowest resonances in the corresponding open system. This excess of
exponentially small decay rates is as dramatic a signature of the
underlying classical behavior as the long wavefunction intensity tails found
in Section~\ref{wis}.

Energy-averaging allows us to sample resonances from all three energy regimes:
wide resonances from near $E=\phi$, very narrow resonances from
near $E=\phi+\pi$, and intermediate-width resonances from
the region described by Eq.~\ref{inters} (the same can be
accomplished by varying a weak magnetic field
at constant energy and thus sweeping through different values of the
phase $\phi$). The total probability to remain in the system
at short times is given by
\begin{equation}
P_{\rm rem}= 1- <S_{\rm lin}> \Gamma t/N = 1- \Gamma_{\rm cl}t
\end{equation}
(where we use the first line of Eq.~\ref{preme} and notice
that $<S_{\rm lin}>=A_{\rm lin}(0)=\langle a|a\rangle =1$ by
normalization). Thus at short times $t \ll \Gamma_{\rm cl}^{-1}$,
the faster-decaying
scarred states and slower-decaying antiscarred states always cancel exactly
to produce a result consistent with our classical expectations. At long
times, we use the second line of Eq.~\ref{preme}
to obtain
\begin{equation}
P_{\rm rem}(t)={<S_{\rm lin}^{-1}> \over \Gamma_{\rm cl} t} \,.
\label{scardec}
\end{equation}
In the strong scarring regime of small $\lambda$, the average 
$<S_{\rm lin}^{-1}>$ is dominated by narrow resonances 
within $O(\lambda)$ of the antiscarring
energy $E=\phi$ (Eq.~\ref{mins}), and gives an exponentially large (in
$\lambda$) enhancement over the predictions of RMT:
\begin{equation}
\label{enh}
<S_{\rm lin}^{-1}> = \left({\lambda \over 2 \pi} \right)^{2}
e^{\pi^2/2 \lambda} \,.
\end{equation}
Thus, the quantum mechanical probability to remain in the system at long times
is very strongly dependent on the location of the opening relative
to the unstable classical orbits of the system. This is in contrast
with the classical long-time probability to remain, which for a small opening
is independent of the location of the opening [the classical distribution
continually being uniformly redistributed over the entire phase space,
on a time
scale very short compared to the decay time of the system]. See
Figs.~\ref{figdecprem},\ref{figdecenh} later in this section for a
numerical example.

Enhancements in the probability to remain at long times are
obtained also for leads centered
slightly away from the periodic orbit; in particular an enhancement
of order $\lambda^2 e^{\pi^2/2 \lambda}$ is found as long as the lead
is within a phase space area
scaling as $\lambda^2 \hbar$ surrounding the orbit. Thus, the long-time
probability to remain in the system averaged over all possible positions
of the lead is given by
\begin{equation}
P_{\rm rem}  = { 1+ O(\hbar \lambda^4 e^{\pi^2/2\lambda}) \over
\Gamma_{\rm cl}t} \,.
\end{equation}
[In principle, contributions from all the periodic
orbits need to be added, however, if orbits with small $\lambda$
exist, they will clearly dominate any such sum.] The result is that
at finite energy, exponentially large (in $1/\lambda$) deviations from
RMT are found even after averaging over the lead position.
In the
$\hbar \to 0$ limit of any given classical system, the RMT
behavior is recovered as the probability of the lead being within
$\hbar$ of a short periodic orbit goes to zero.

Classically, not only is the long-time probability to remain independent
of the lead position, but it is also largely independent of the
initial probability distribution in the system (as long as this 
distribution is not concentrated in a very narrow
corridor starting at the lead and having width scaling as the lead size).
Similarly, the remaining classical probability distribution at long
times is always evenly distributed in the classical phase space, again
with the exception of a similar narrow corridor. Both of these statements
are modified by quantum mechanical effects associated with the
unstable periodic orbits of the system~\cite{open}. In the non-overlapping
resonance regime of small $\Gamma$, the long-time probability
to remain in state $|b\rangle$ given a decay channel at $|a\rangle$, as well
as the total probability to remain in the system given an initial
probability concentrated at $|a\rangle$ are {\it both} given by
\begin{equation}
\label{premb}
P_{\rm rem}^b(t) = \sum_n |\langle b|n\rangle|^2 e^{-\Gamma_n t} \,.
\end{equation}
If $|b\rangle$ is not located on the same short periodic orbit as 
$|a\rangle$ (and not related to it by symmetry), the intensities 
$|\langle b|n\rangle|^2$ follow a chi-squared distribution
independent of the decay rates $\Gamma_n\sim |\langle a|n\rangle|^2$,
and the mean value of this intensity at a given energy scales as
$S_{\rm lin}^b$, the smoothed LDOS at $|b\rangle$. Then at energy $E$ we
have
\begin{equation}
P_{\rm rem}^b =
{S_{\rm lin}^b(E) \over 1 + S^a_{\rm lin}(E) \Gamma t/N} \nonumber \to
{S_{\rm lin}^b(E) \over S^a_{\rm lin}(E)}  { 1 \over \Gamma_{\rm cl}t} \,.
\end{equation}
Again averaging over energy we obtain the long-time ratio of the remaining
probability density at $|b\rangle$ compared to the average remaining
probability density at long times:
\begin{equation}
\label{ratio}
{ P_{\rm rem}^b \over P_{\rm rem} } =  {< S_{\rm lin}^b / S_{\rm lin}^a >
\over  < 1 /S_{\rm lin}^a > }\,.
\end{equation}

The ratio obtained above
strongly deviates from unity if both $|a\rangle$ and $|b\rangle$
are located on short unstable periodic orbits, so that their linearized
spectral envelopes $S_{\rm lin}^b$ and $S_{\rm lin}^a$ are nontrivial.
Consider two specific cases. If $S_{\rm lin}^a$ and $S_{\rm lin}^b$
are located on (distinct) periodic orbits of the same classical period,
action, and stability exponent $\lambda$, then
the two envelopes are identical and we obtain
the exponential suppression
\begin{equation}
{ P_{\rm rem}^b \over P_{\rm rem} } = <(S_{\rm lin}^a)^{-1}>^{-1}
\sim \lambda^{-2} e^{-\pi^2/2\lambda}\,.
\end{equation}
On the other hand, if the two spectral envelopes are exactly out of phase
in the region of interest, i.e. $S^b_{\rm lin}(E)=S^a_{\rm lin}(E+\pi)$,
the ratio is dominated by the peak in $S_{\rm lin}^b$, and becomes
an {\it enhancement} factor $\lambda^{-1}$ for small $\lambda$.
We note that this suppression or enhancement of probability on orbits {\it
other} than the one on which the lead is located is of course a quantum
interference effect, having no counterpart in the classical dynamics
of the system. Even stronger interference effects are observed on the orbit
where the lead is located, due to the very strong correlations
(found in Section~\ref{ims}) between densities of states at different points
on the same orbit; see \cite{open} for details.

The data in this section is taken from \cite{open}, and
was originally obtained using an ensemble of
perturbed cat maps. See \cite{open} for a fuller discussion of the numerical
evidence. In Fig.~\ref{figdecprem} is shown the measured probability to remain
in the system as a function of scaled time $t'=\Gamma_{\rm cl}t$, for a generic
lead location (plusses) and for a lead placed on an unstable
periodic orbit with exponent $\lambda=0.96$ (squares). The classical
exponential decay law (dotted curve) is shown for comparison. We see
that the numerical results for a generically placed lead
agree well with the RMT prediction of Eq.~\ref{rmtdec}, shown as a dashed line.
The long-time probability to remain given a lead placed on the periodic orbit
shows a large enhancement factor, consistent with the theoretical
prediction of Eq.~\ref{scardec} (solid line). For this value of $\lambda$,
the predicted enhancement factor at long times is $11.0$. The numerical
data is obtained for a system size $N=120$, and the decay rate per
step in the exit channel is $\Gamma=0.1$.

In Fig.~\ref{figdecenh}, the predicted enhancement factor $<S_{\rm lin}^{-1}>$
in the long-time probability to remain is plotted as a solid curve, as
a function of the instability exponent $\lambda$ of the orbit on which the lead
is located. The prediction is valid in the semiclassical limit $N \to \infty$;
numerical data is presented for $N=120$ (plusses) and $N=240$ (squares). We
notice the very large enhancement factors that are obtained even for moderate
values of $\lambda$, in accordance with the exponential asymptotic behavior
of Eq.~\ref{enh}. Finally, in Fig.~\ref{figdecrat} we give an example
of the enhancement and suppression of intensity at long times found on an
orbit other than the one on which the lead is located, as the relative
phase between the two orbits is varied. In practice, a curve of this kind
can be obtained by varying a magnetic flux line or a weak magnetic field
enclosed by one of the orbits. The theoretical prediction
(Eq.~\ref{ratio}) and numerical
data are obtained for the case where both orbits have instability 
$\lambda=1.76$. Because the stretching factor $e^\lambda$ for one iteration
of the orbit is quite large, both the
intensity 
suppression when the orbits are in phase, and the enhancement when
the two orbits are out of phase are quite moderate in this example.

An application of scarring methods to the study of conductance peaks
in chaotic system with two leads is
upcoming~\cite{cond}; here we summarize the ideas and the major results.
We again consider the regime where the leads are narrow (and the temperature
low), so that the resonance peak width is small compared to the level
spacing. The height of the $n$-th conductance peak
is then given in the appropriate units by
\begin{equation}
G \sim {\Gamma_{na} \Gamma_{nb} \over \Gamma_{na} + \Gamma_{nb}} \,,
\end{equation}
where $\Gamma_{na}$, $\Gamma_{nb}$ are the decay rates of the $n$-th
resonance through each of the two leads. If the openings are of the same
size, this reduces to
\begin{equation}
G \sim {|\langle n|a \rangle|^2 |\langle n|b \rangle|^2 \over
|\langle n|a \rangle|^2 + |\langle n|b \rangle|^2} \,,
\end{equation}
where we have taken out an overall constant proportional to the
size of the opening. In RMT, the two intensities $|\langle n|a \rangle|^2$
and $|\langle n|b \rangle|^2$ are of course taken to be independent
chi-squared variables (of one or two degrees of freedom, for real
or complex overlaps, respectively), producing the desired distribution
of peak heights. Several situations are now considered which show marked
deviations from the RMT predictions.

(i) One of the leads, say $a$, is optimally placed
on a short unstable periodic orbit, of
exponent $\lambda$.  Then conductance peak heights show periodic fluctuations
in energy, with larger conductances at the scarring energies, and smaller
conductances far from these energies. In particular, for small $\lambda$
the mean value of $|\langle n|a \rangle|^2$ becomes large (as $\lambda^{-1}$)
at the peak scarring energies, and the peak heights are then governed entirely
by the fluctuations in $|\langle n|b \rangle|^2$. Near the minimum of
$S_{\rm lin}^a$, the effect is more dramatic: the peak height
again becomes a simple chi-squared variable, here governed entirely
by fluctuations in $|\langle n|a \rangle|^2$, but the mean value of the peak
height is exponentially small in $\lambda$ (see Eq.~\ref{mins}). A fraction
of $O(\lambda)$ of all peaks will have the suppression of Eq.~\ref{mins}; this
excess of exponentially small conductance peaks will of course persist after
averaging over either energy or weak magnetic field strength.

(ii) The two leads are placed on two orbits whose spectral envelopes
are ``in phase" in the energy range of interest. Say for simplicity
that the two orbits also have the same exponent $\lambda$, so
$S_{\rm lin}^a=S_{\rm lin}^b$. At each energy, the conductance peak heights
now follow the same distribution as in RMT, but the mean value is
proportional to $S_{\rm lin}(E)$. Thus, at the scarring energies the
conductance peak heights are increased by $O(\lambda^{-1})$ over the mean
RMT value, whereas at the antiscarring energies the heights
are suppressed by $O(\lambda^2 e^{-\pi^2/2\lambda})$. The energy-averaged
(or magnetic-field averaged) distribution can also be obtained, and shows
an enhancement in probability for large and small peak heights (compare
the tail of the wavefunction intensity distribution in Section~\ref{wis}).

(iii) The two orbits are ``out of phase". Then strong suppression of
conductance peak heights is predicted everywhere, and the mean peak height
is also exponentially small in $\lambda$. The largest peak heights
appear at energies intermediate between the maxima of
$S_{\rm lin}^a$ and $S_{\rm lin}^b$, where $S_{\rm lin}^a \approx
S_{\rm lin}^b$. Even at these energies, the conductance is exponentially
suppressed compared to the RMT prediction.

(iv) The two leads are located on the same orbit, or on two orbits related
by a symmetry. Here $|\langle n|a \rangle|^2=|\langle n|b \rangle|^2$,
and the peak height distribution is again given by a simple
chi-squared distribution, with mean given by $S_{\rm lin}(E)$. The
energy-averaged mean peak height is independent of the stability exponent
$\lambda$, but once again we predict power-law (in $\lambda$)
enhancement of the
conductance near the scarring energies, and exponential suppression near the
antiscarring energies.

\section{Conclusion}
\label{conc}

We have seen that short unstable periodic orbits
leave a strong imprint on the long-time and stationary properties of
a quantum chaotic system, even though the corresponding classical dynamics
loses all memory of these structures at long times. This insight can be
extended to other situations where short-time classical structures
lead to modifications of RMT predictions concerning the structure
of quantum wavefunctions. Sundaram and Scharf~\cite{sundaram} study
periodic orbits in {\it complex} phase space, with complex action,
and show that these ``ghost" orbits  can also produce
wavefunction scars. 

All of the discussion in the present paper applies to isolated orbits
in a system of any dimension, though specific formulas need to be
adjusted to take proper account of multi-dimensional Gaussian
wavepackets and their short-time evolution. In higher-dimensional
chaotic systems, classically invariant manifolds containing large
families of individual orbits can also lead to scar-like effects,
if the classical rate of escape away from these manifolds is slow.
Examples include highly excited eigenstates in three-dimensional
billiards~\cite{prosen} and wavefunctions of few-body chaotic systems
made up of identical particles with rotation and permutation
symmetries~\cite{papenbrock}.

Vilela Mendes has analyzed ``saddle scars," structures which arise
from unstable harmonic motion along the stable manifold of a
classical saddle point~\cite{vilelamendes}. Blumel et al.
extend quantum chaos techniques, including analysis of scars, 
to problems in which ray splitting surfaces are present~\cite{blumel}.
Delande
and Sornette study acoustic radiation from membranes, and find
localization in the radiation directivity that can be attributed to 
wavefunction scars~\cite{delande}

The insights of scar theory can be extended also to short-time
effects not associated with any classical structures. These
short-time recurrences may result from diffractive scattering
or other hard quantum dynamics far from the semiclassical regime.
Thus, in a diffusive system, the local, short-time behavior of a
quantum particle may lead in certain models to strong effects
on the statistics of anomalous wavefunctions, effects not captured
by the standard field theory methods~\cite{diffwis} which generally
ignore dynamics on scales shorter than a mean free path. The scar
picture may prove to be a unifying framework for studying anomalous
localization on graphs, lattices, in disordered systems, and in
S-matrix eigenstates. These extensions of scar ideas to non-classical
short-time behavior are now under development.

\section{Acknowledgements}

This research was supported by the National Science Foundation under
Grant No. 66-701-7557-2-30.

\begin{figure}
\caption{
In (a), the full spectrum is plotted along with the
linear envelope (dotted line), an intermediate envelope corresponding to
$|T|<30$ (solid line), and a semiclassical intermediate envelope (dashed
line). In (b), a portion of the spectrum (solid) is compared to the
spectrum obtained using semiclassical eigenstates (dashed). In (c),
the quantum spectrum has been divided out by the linear envelope of (a)
(after~\protect\cite{nlscar}).
}
\label{fignumspec}
\end{figure}

\begin{figure}
\caption{
A plot of the actual value of the inverse participation
ratio (IPR) for a wavepacket centered on a periodic orbit (squares),
{\it versus}
the value predicted by the linear theory. IPR's in the semiclassical
approximation are plotted using the `+' symbol, and IPR's in a baker
map with random matrix theory nonlinear behavior are plotted with triangles
(after~\protect\cite{nlscar}).
}
\label{figiprvslin}
\end{figure}

\begin{figure}
\caption{
The two-point spectral correlation function of the
scaled spectrum (diamonds) is compared to the correlation function of the
raw spectrum (plusses), after ensemble and energy averaging
(after~\protect\cite{nlscar}).
}
\label{fig2ptcorrel}
\end{figure}

\begin{figure}
\caption{
Cumulative
wavefunction intensity distribution (a) as measured by a test state
centered on a periodic orbit with instability $\lambda=\log 2$, plotted as
the upper thick curve with
scarring theory prediction given by dashed curve, and (b) averaged over the
entire phase space of size $200h$, plotted as lower thick curve
with theory given by solid curve. The dotted
line is the Porter-Thomas law
(after~\protect\cite{sscar}).
}
\label{figwis1}
\end{figure}

\begin{figure}
\caption{
Cumulative wavefunction intensity distribution after ensemble
averaging over systems with classical orbits of different
instability exponents. The scar theory tail is given by the solid line,
and the dotted curve is the RMT prediction
(after~\protect\cite{sscar}).
}
\label{figwis2}
\end{figure}

\begin{figure}
\caption{
The correlation $C_{ab}$ (Eq.~\ref{cabdef})
between local densities of states for two
wavepackets $|a\rangle$ and $|b\rangle$, lying on the same periodic
orbit of instability exponent $\lambda$, is plotted as a function of
width parameter $z$.
The upper and lower curves correspond to $\lambda=\log 5$ and
$\lambda=\log 10$, respectively.
The width (along the unstable manifold) of
wavepacket $|b\rangle$ is $e^{\lambda(n+z)}$ times that of $|a\rangle$,
where $n$ is an integer 
(after~\protect\cite{is}).
}
\label{figiscab}
\end{figure}

\begin{figure}
\caption{
The inverse participation ratio (IPR) for hyperbolic test state
$|\Psi\rangle$ is plotted versus the log-time cutoff $T_0$ (see
Eqs.~\ref{cohdefth},~\ref{t0def}), for various values of system size
$N$. From bottom to top, the five curves correspond to $N=50$, $100$, $200$,
$400$, and $800$. For each $N$, $26$ points are plotted, for $c=(1.1)^j$,
$j=-20 \ldots +5$. The orbit has exponent $\lambda = |\log 0.18|$.
The upper dashed curve is the $N \to \infty$
theoretical prediction, which converges to the
asymptotic
prediction of Eq.~\ref{iprpsi2} (lower dashed line) for large $T_0$. The
linearized theory is expected to start breaking down for $c \ge 1$ (rightmost
six points on each data curve)
(after~\protect\cite{is}).
}
\label{figisipr}
\end{figure}

\begin{figure}
\caption{
Smoothed local densities of states are plotted for the off-resonance
universal hyperbolic
test state $|\Psi\rangle$, for off-resonance angle $\theta$ ranging from $0$
(tallest peak) in steps of $\pi/10$ (to the left), through $-2\pi$,
on a periodic orbit with exponent
$\lambda = |\log 0.18|$.
The system size is $N=800$, and the cut-off constant $c$ is set to $0.6$
(see Eq.~\ref{t0def}) (after~\protect\cite{is}).
}
\label{figisth}
\end{figure}

\begin{figure}
\caption{
Husimi plots of
the universal test state $\Psi$ for off-resonance parameter values
(a) $\theta/\lambda=0.8$ and (b) $\theta/\lambda=2.5$. The
linearizable region is taken to be much larger than the displayed area
of size  $12 \sqrt\hbar \times 12 \sqrt\hbar$, centered on the
periodic orbit (after~\protect\cite{is}).
}
\label{figishus}
\end{figure}

\begin{figure}
\caption{
Smoothed local densities of states for a Gaussian
wavepacket placed on a period-$2$ orbit (double-peaked solid curve),
and for the universal test state $\Psi$ constructed on the same orbit
(tall single peak). The dashed curves represent theoretical
predictions based on the linearized dynamics near the periodic
orbit in question. The periodic orbit
has a total exponent $\lambda=|\log 0.168|$ over the two-step period
(after~\protect\cite{is}).
}
\label{figistp2}
\end{figure}

\begin{figure}
\caption{
Smoothed local densities of states $S_{\rm lin}(E)$ are plotted as
a function of energy on a periodic orbit of instability exponent $\lambda=0.20$
(solid curve) and on an orbit with $\lambda=0.15$ (dashed curve).
We observe
the peak at the EBK quantization energy $E=0$ which scales
as $\lambda^{-1}$, the exponential decay between $E=0$ and $E=\pi$
(Eq.~\ref{inters}), and the minimum at the anti-EBK energy $E=\pi$,
which is exponentially small in $\lambda$
(Eq.~\ref{mins}). The RMT prediction
$S_{\rm lin}(E)=1$, which is applicable away from any short periodic orbit,
is plotted as a dotted line (after~\protect\cite{open}).
}
\label{figdecslin}
\end{figure}

\begin{figure}
\caption{
The probability to remain in the open quantum system is
plotted as a function of scaled time $t'=\Gamma_{\rm cl}t$. The classical
prediction $\exp (-t')$ is shown as a dotted curve. The quantum probability
to remain for a generic lead location (plusses) compares well with the
RMT prediction $1/(1+t')$ (dashed curve). For a lead placed on a short
periodic orbit with instability exponent $\lambda=0.96$, we obtain the enhanced
long-time probability to remain (squares), which agrees with the scar
theory prediction $<S_{\rm lin}^{-1}>/t'$ (solid line). The system size used
for obtaining the data is $N=120$ and the decay rate per step
in the exit channel is $\Gamma=0.1$ (after~\protect\cite{open}).
}
\label{figdecprem}
\end{figure}

\begin{figure}
\caption{
The long-time enhancement factor of the probability to remain
in a system when the lead is placed on a periodic orbit is plotted
as a function of the instability exponent of the orbit. Data is shown
for $N=120$ (plusses) and $N=240$ (squares).
The $N \to \infty$ theoretical
prediction $<S_{\rm lin}^{-1}>$ is shown as a solid curve. We see the
exponential increase in the probability to remain as the exponent $\lambda$
decreases (the $\lambda \to 0$ asymptotic form is given in Eq.~\ref{enh}).
For large $\lambda$, the enhancement factor converges to $1$, the RMT
prediction (after~\protect\cite{open}).
}
\label{figdecenh}
\end{figure}

\begin{figure}
\caption{
The predicted relative intensity at long times on a periodic orbit other
than the one containing the lead is plotted as a function of the relative
phase between the two orbits (solid curve). For reference, the phase-space
averaged intensity is plotted as a dashed line. Intensity suppression
is predicted and observed when the orbits are in phase, and enhancement
is seen when the orbits are exactly out of phase. 
The data is collected for an ensemble of systems with
$N=80$, and each orbit has instability exponent $\lambda=1.76$
(after~\protect\cite{open}).
}
\label{figdecrat}
\end{figure}

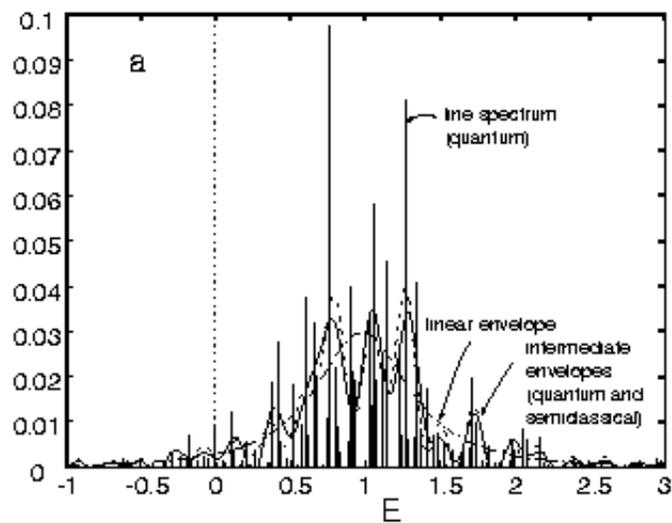
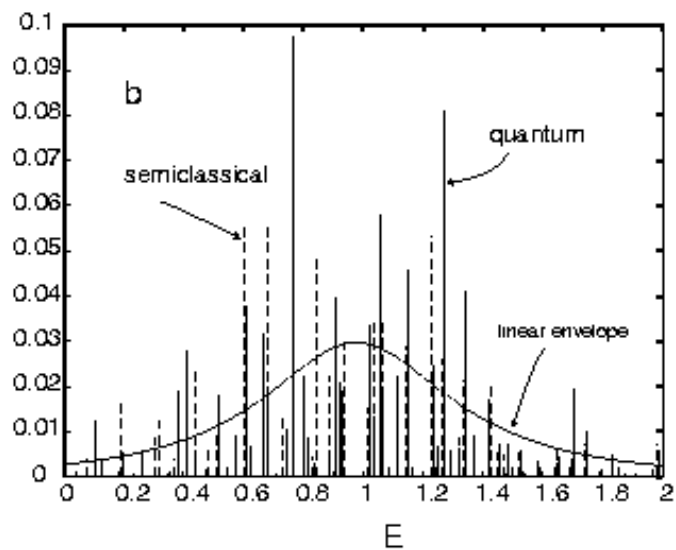
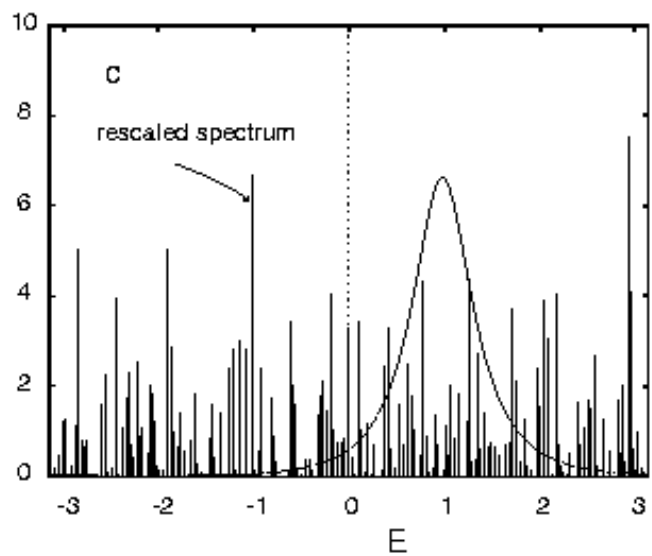
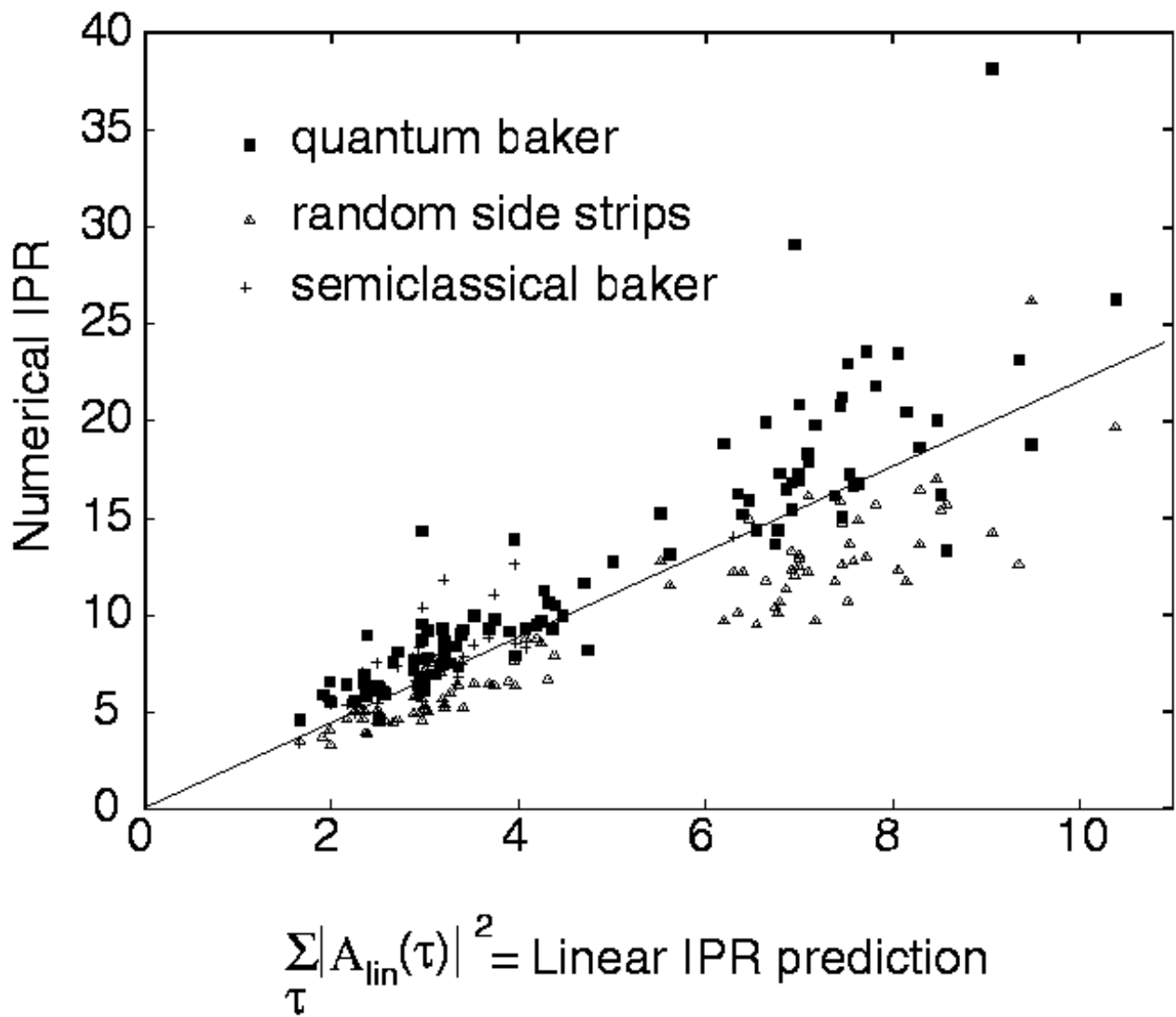
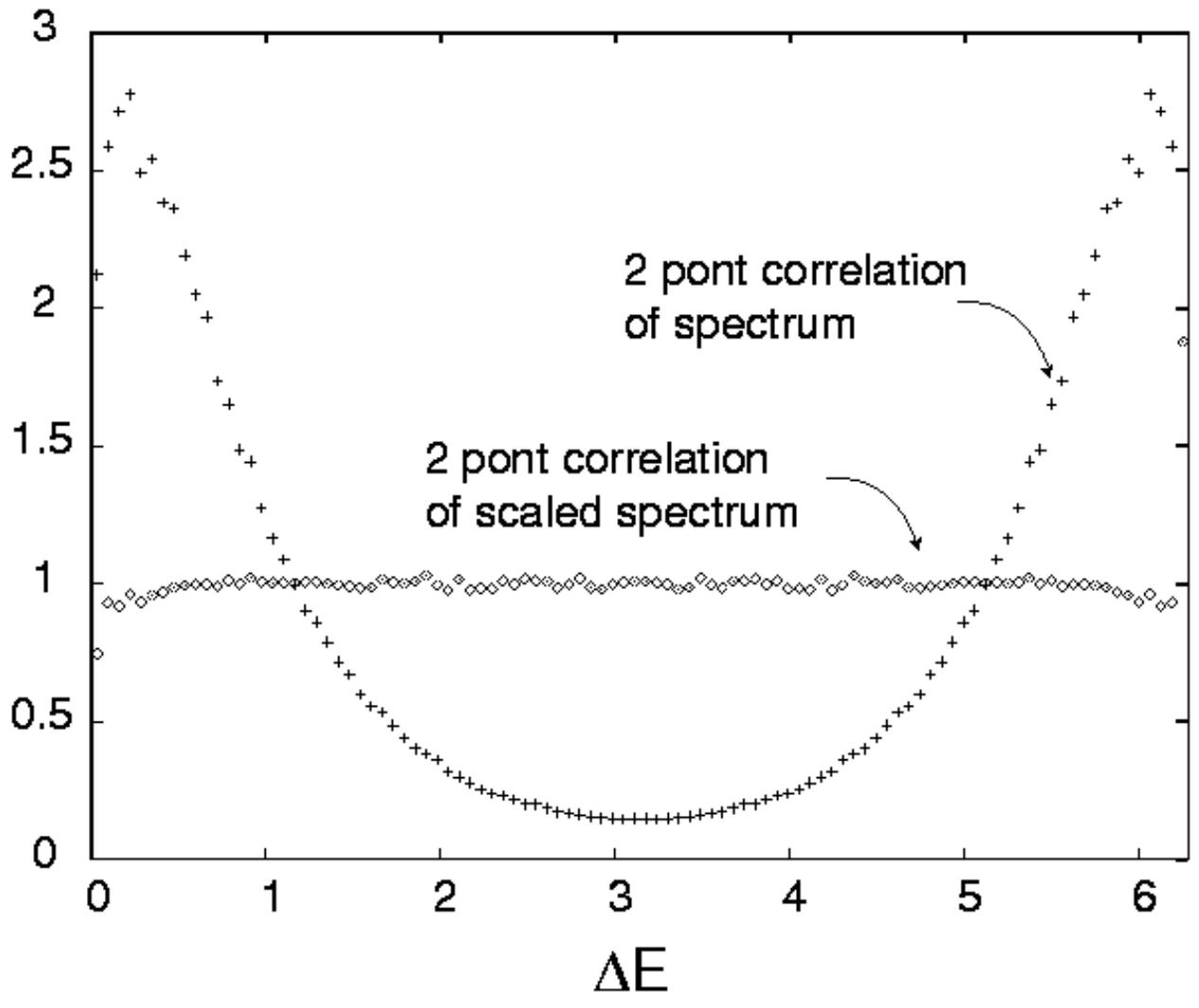
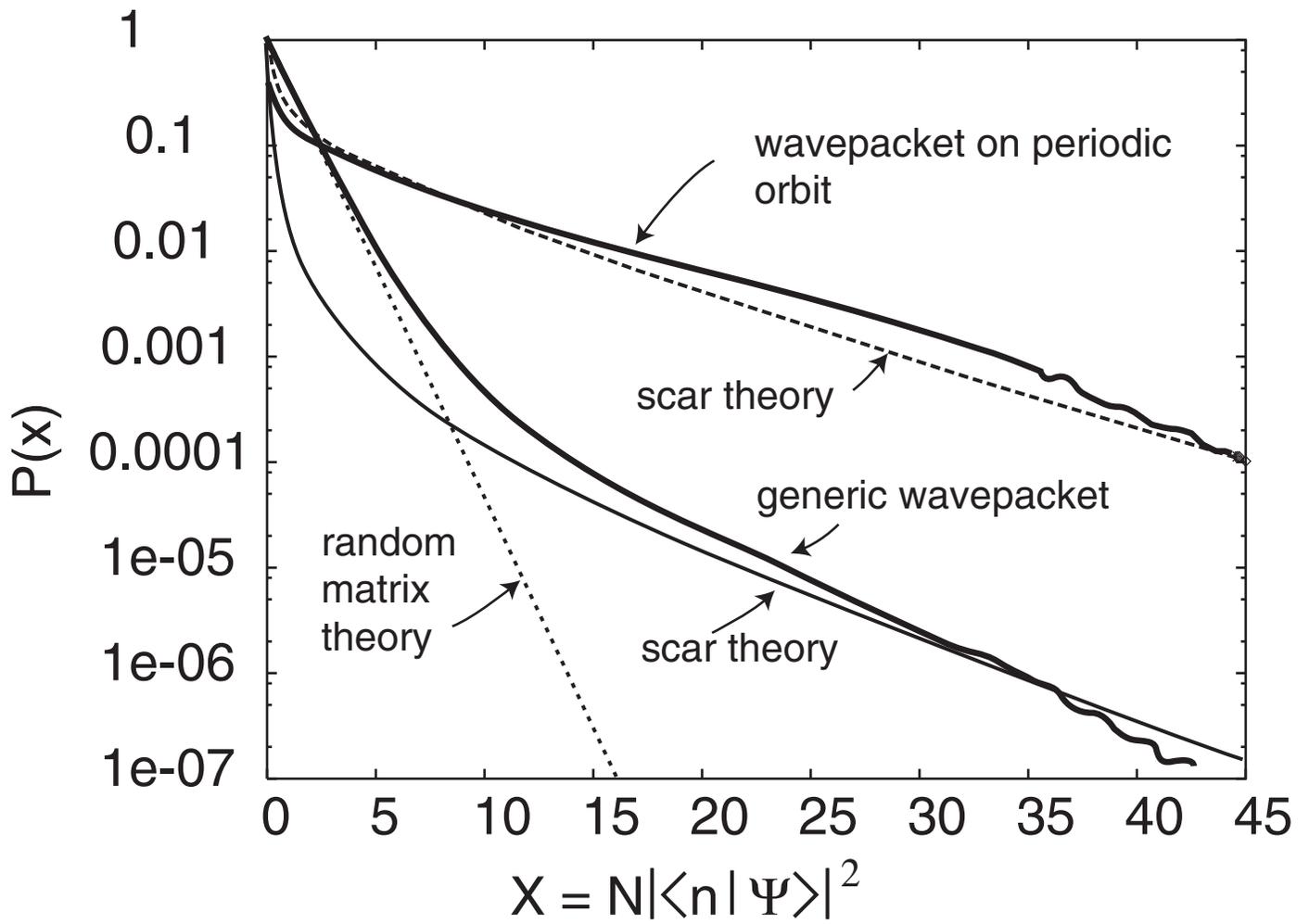
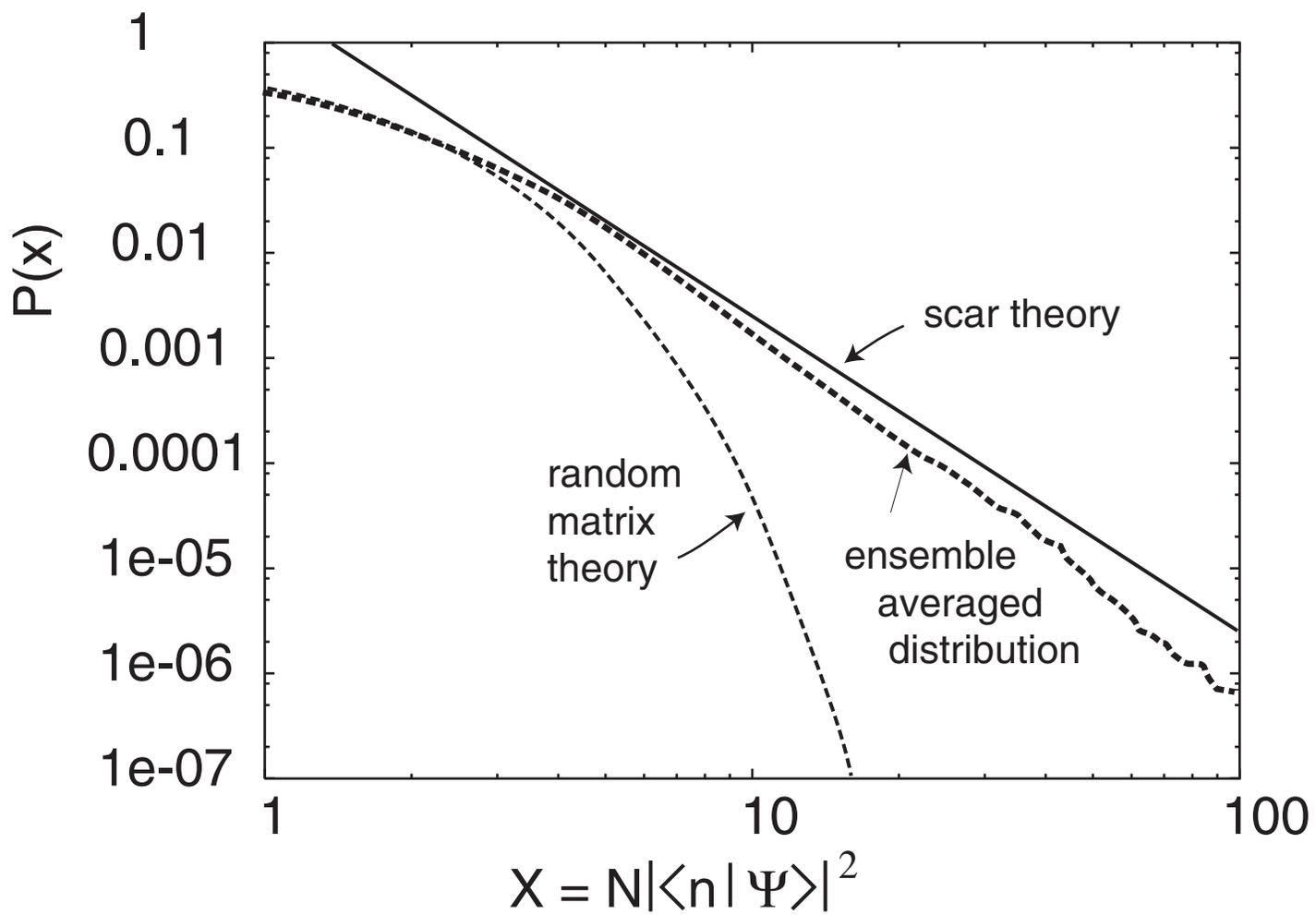
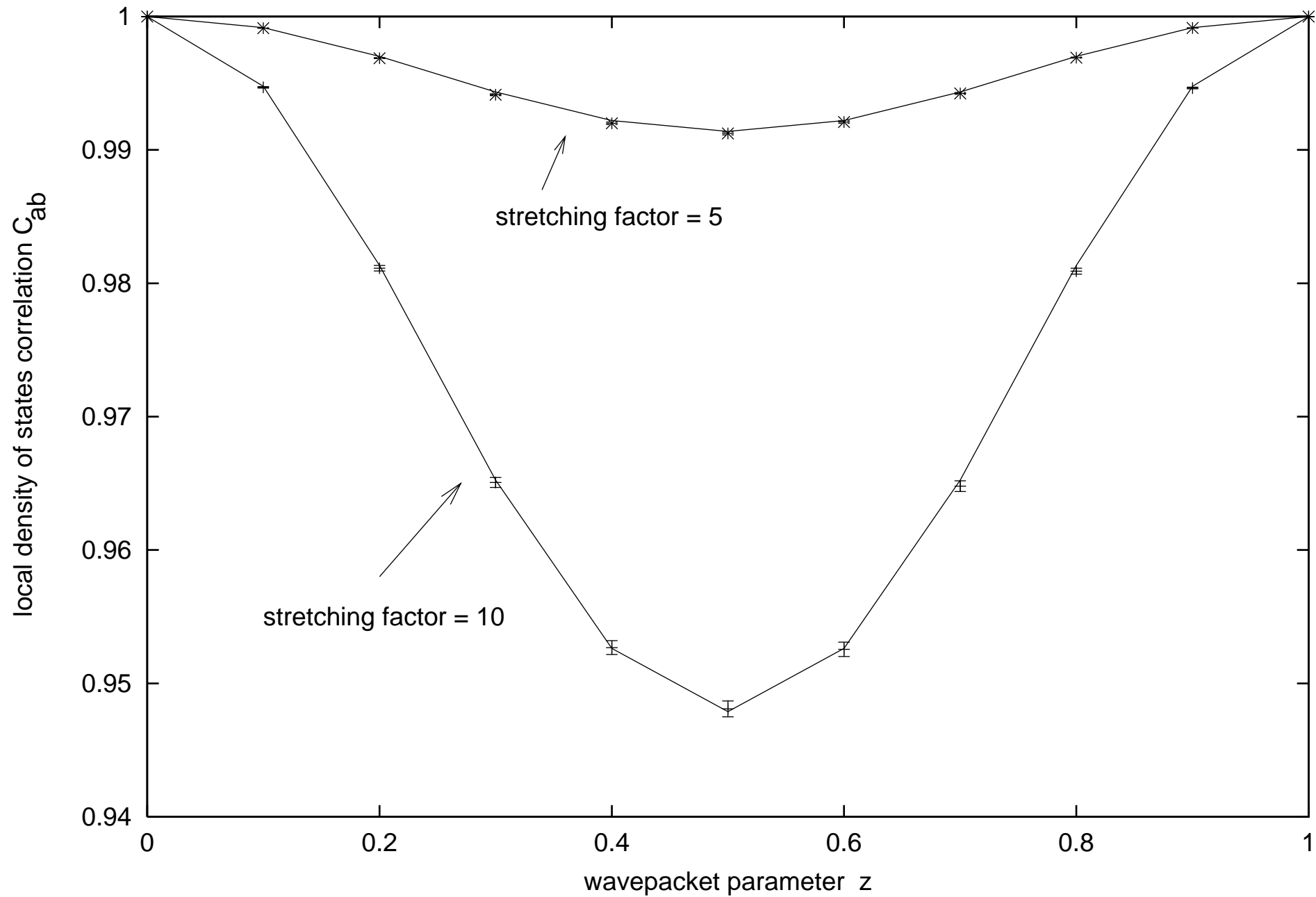
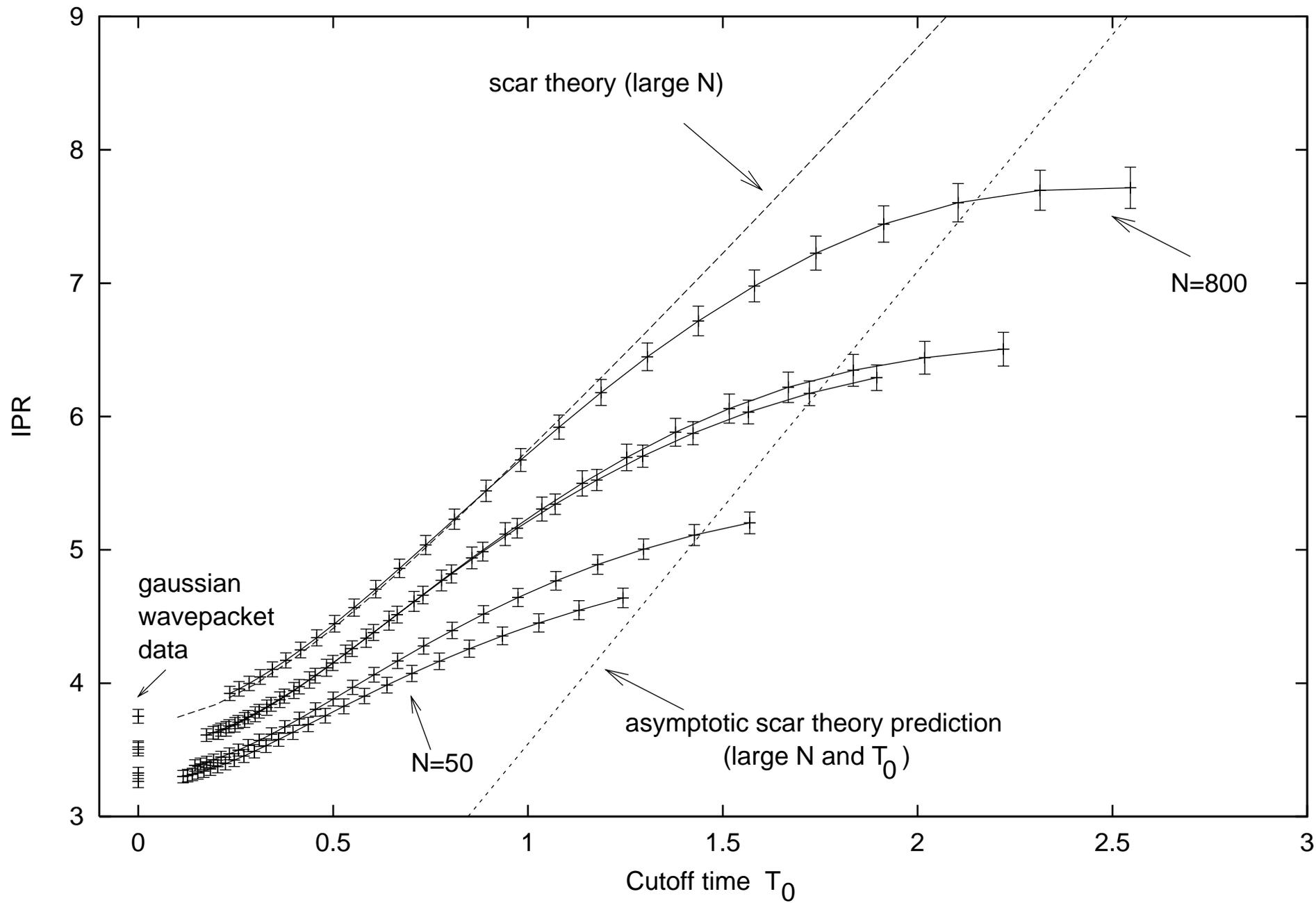
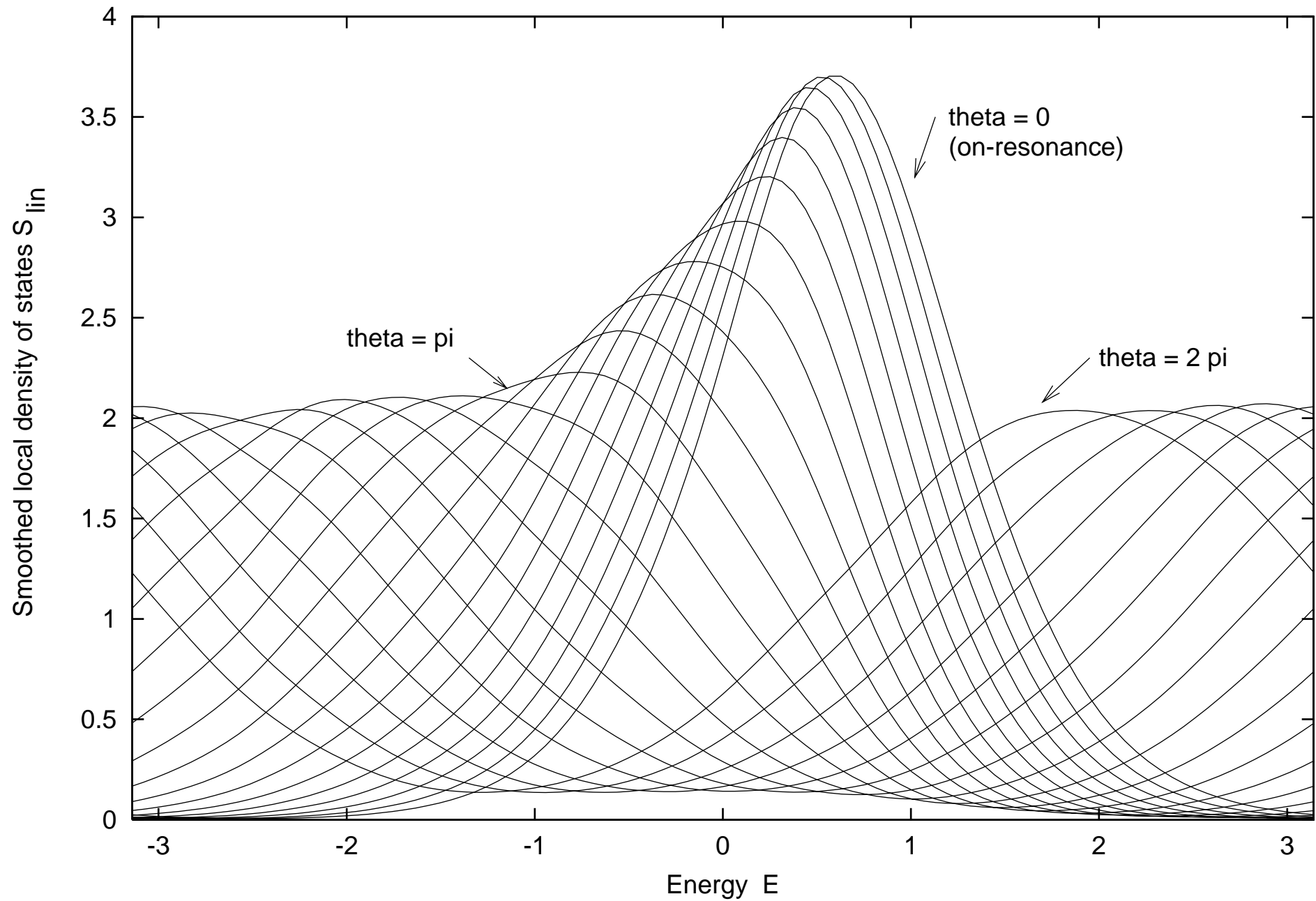
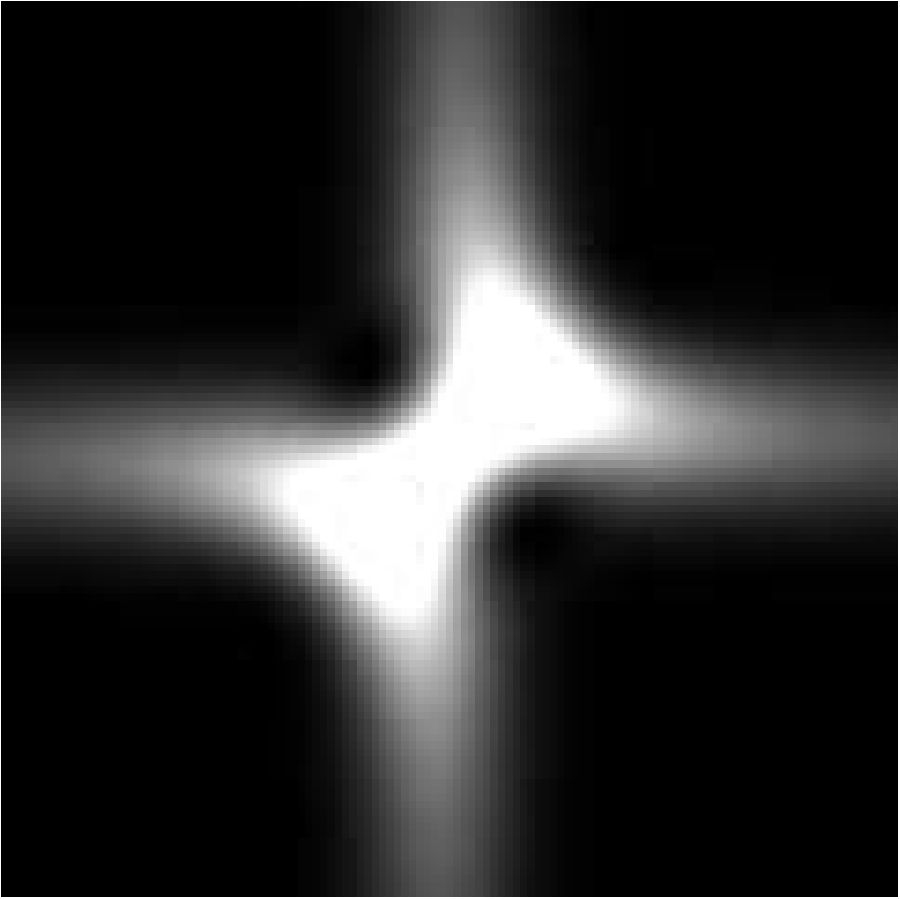
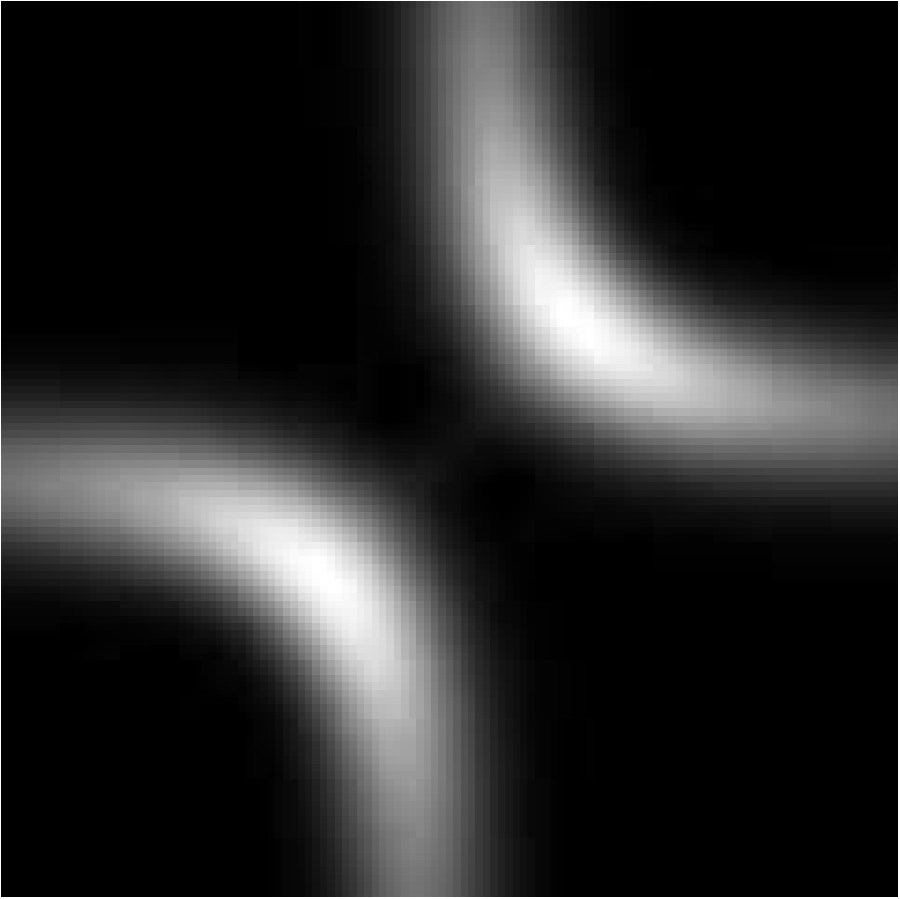
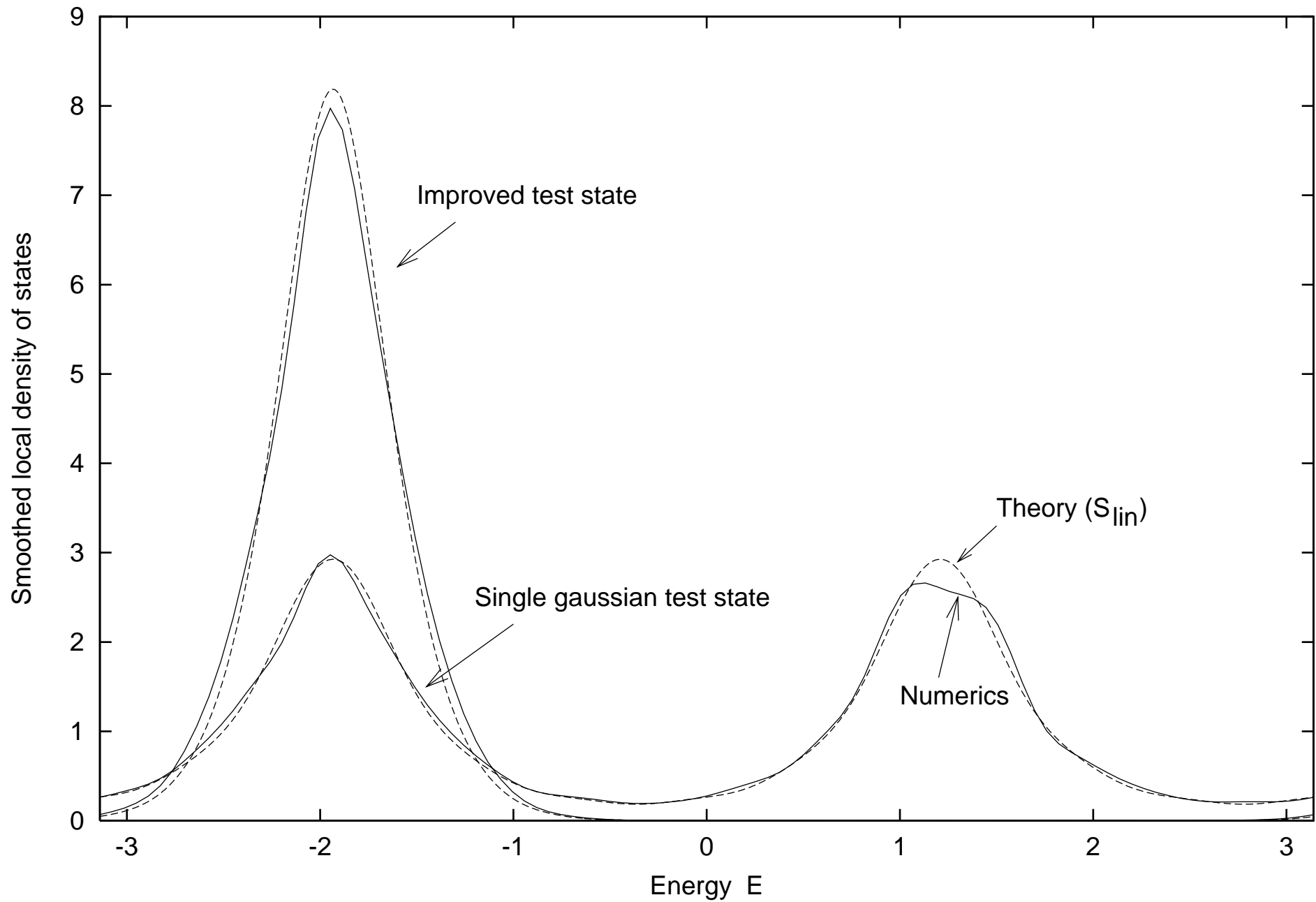
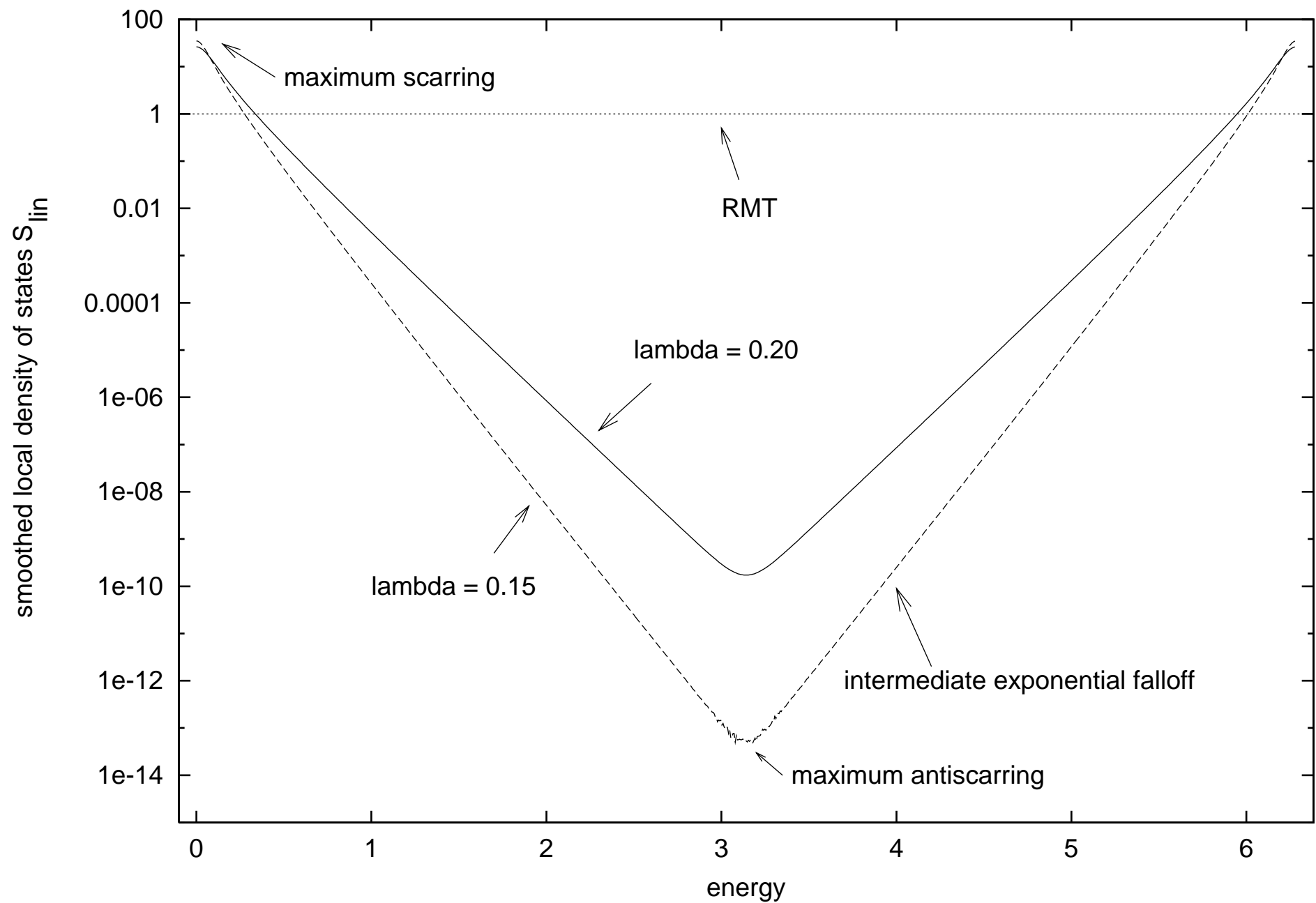
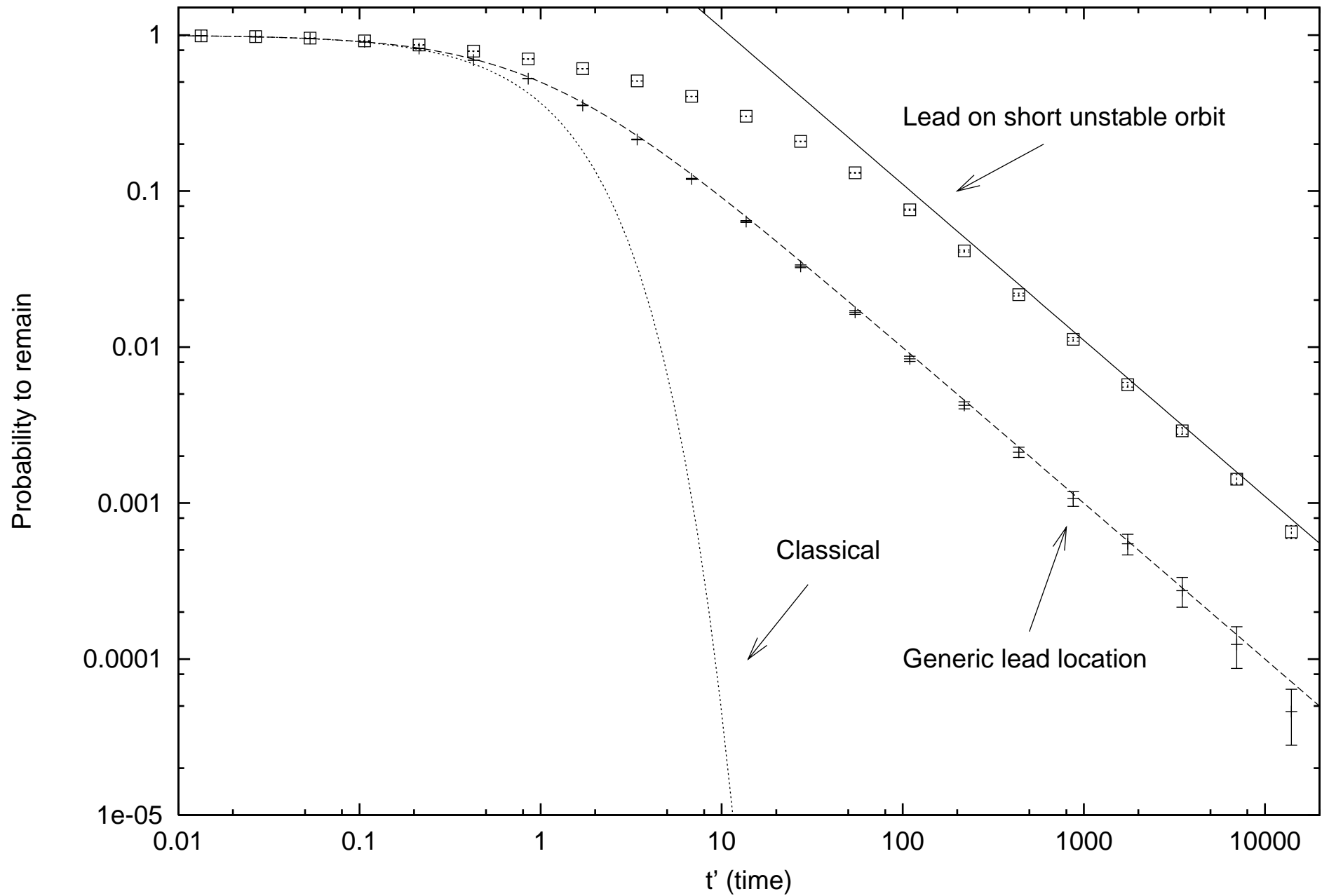
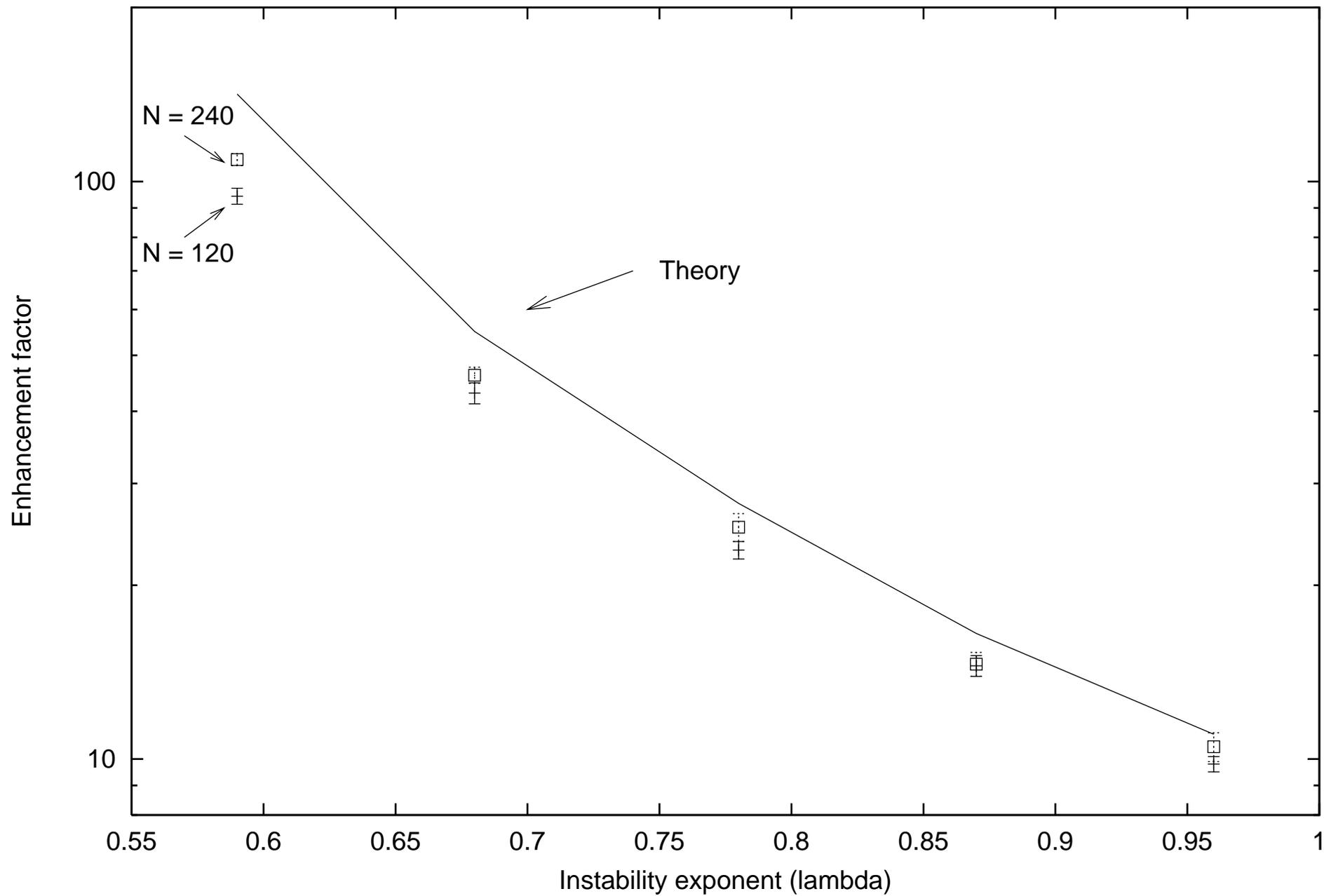
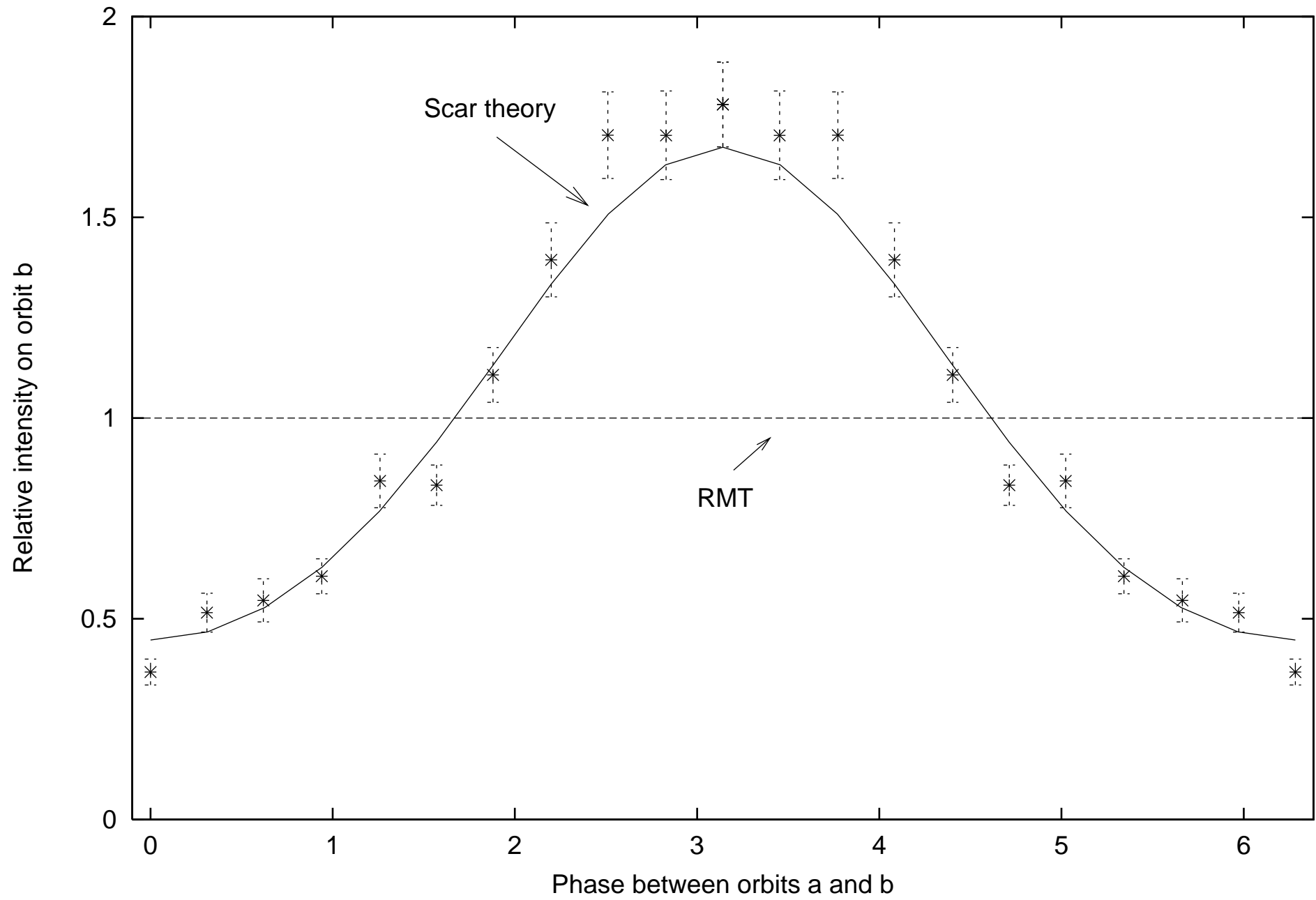
\end{document}